\documentclass[aps,prb,floatfix,twocolumn,nobibnotes,longbibliography,superscriptaddress]{revtex4-2}
\usepackage{graphicx}
\usepackage{todonotes}
\usepackage{amsmath}
\usepackage{amsmath}
\usepackage{upgreek}
\usepackage{color}
\usepackage{standalone}
\usepackage{pgfplots}
\usetikzlibrary{positioning}
\usepackage{array}
\newcolumntype{C}[1]{>{\centering\let\newline\\\arraybackslash\hspace{0pt}}m{#1}}
\usepackage{multirow}
\usepackage{hyperref}
\usepackage{float}
\usepackage{todonotes} %Am Ende entfernen!!!
\usepackage{rotating}
\usepackage{gensymb}\usepackage{xcolor}

\usepackage{lmodern}        % Latin Modern
\begin{document}

\newcommand{\dd}[1]{\frac{\text{d}}{\text{d}#1}}
\newcommand{\de}{\text{d}}
\newcommand{\chipe}{ \chi^m_{\perp} }
%https://www.overleaf.com/project/5f113ed97499b400015af1a3\newcommand{\chipa}{ \chi^m_{\parallel} }
\renewcommand{\phi}{ \varphi }
\newcommand{\unitvec}[1]{\mathbf{\hat #1}}
\newcommand{\ve}[1]{{\mathbf{#1}}}
\newcommand{\degC}[1]{{$^{\rm\circ}$}}
\newcommand{\cm}  {$\rm cm^{-1}$}
%\sloppy
\title{On the Issue of Textured Crystallization of Ba(NO$_3$)$_2$ in\\ Mesoporous SiO$_2$: Raman Spectroscopy and Lattice Dynamics Analysis}

\author{Yaroslav Shchur}
%\corref{Shchur}*
\affiliation{Institute for Condensed Matter Physics, 1 Svientsitskii str., 79011, Lviv, Ukraine}
\email{shchur@icmp.lviv.ua}
\author{Guillermo Beltramo}
\affiliation{Institute of Biological Information Processing Mechanobiology (IBI-2), Forschungszentrum Juelich, D-52425 Juelich, Germany}
\author{Anatolii S. Andrushchak}
\affiliation{Lviv National Polytechnic University, 12 S. Bandery str., 79013, Lviv, Ukraine}
\author{Svetlana Vitusevich}
\affiliation{Institute of Bioelectronics (IBI-3), Forschungszentrum Juelich, D-52425 Juelich, Germany}
\author{Patrick Huber}
\affiliation{Hamburg University of Technology, Institute for Materials and X-Ray Physics, Denickestr. 10, 21073 Hamburg, Germany}
\affiliation{Centre for X-Ray and Nano Science CXNS, Deutsches Elektronen-Synchrotron DESY, Notkestr. 85, 22603 Hamburg, Germany}
\affiliation{Centre for Hybrid Nanostructures CHyN, Hamburg University, 22607 Hamburg, Germany}
\author{Volodymyr Adamiv}
\affiliation{O.G. Vlokh Institute of Physical Optics, 23 Dragomanova str., 79005, Lviv, Ukraine}
\author{Ihor Teslyuk}
\affiliation{O.G. Vlokh Institute of Physical Optics, 23 Dragomanova str., 79005, Lviv, Ukraine}
\author{Nazarii Boichuk}
\affiliation{Institute of Bioelectronics (IBI-3), Forschungszentrum Juelich, D-52425 Juelich, Germany}
\author{Andriy V. Kityk}
\affiliation{Faculty of the Electrical Engineering, Czestochowa University of Technology, Al. Armii Krajowej 17, 42-200, Czestochowa, Poland}

\date{\today}

\begin{abstract}
The lattice dynamics of preferentially aligned nanocrystals formed upon drying of aqueous Ba(NO$_3$)$_2$ solutions in a mesoporous silica glass traversed by tubular pores of approximately 12 nm are explored by Raman scattering. To interpret the experiments on the confined nanocrystals polarized Raman spectra of bulk single crystals and X-ray diffraction experiments are also performed. Since a cubic symmetry is inherent to Ba(NO$_3$)$_2$, a special Raman scattering geometry was utilized to separate the phonon modes of A$_g$ and E$_g$ species. Combining group-theory analysis and \textit{ab initio} lattice dynamics calculations a full interpretation of all Raman lines of the bulk single crystal is achieved. Apart from a small confinement-induced line broadening, the peak positions and normalized peak intensities of the Raman spectra of the nanoconfined and macroscopic crystals are identical. Interestingly, the Raman scattering experiment indicates the existence of comparatively large, $\sim$10-20 $\mu$m, single-crystalline regions of Ba(NO$_3$)$_2$ embedded in the porous host, near three orders of magnitude larger than the average size of single nanopores. This is contrast to the initial assumption of non-interconnected pores. It rather indicates an inter-pore propagation of the crystallization front, presumably via microporosity in the pore walls.

%keywords: Raman scattering, Ba(NO$_3$)$_2$, porous SiO$_2$, phonon, density functional theory, confinement effect, X-ray diffraction

%Ba(NO$_3$)$_2$ nanocrystals are synthesized in mesoporous silica matrix.

% Short title
%\shorttitle{Short title: Ba(NO$_3$)$_2$ crystals grown in porous SiO$_2$ matrix}

% Short author
%\shortauthors{Short authors: Ya. Shchur, G. Beltramo, A.S. Andrushchak \textit{et al}}

%\highligths{Highligths:

%	1. Raman spectroscopy surface mapping is used for structure probing of the newly synthesized crystal.
	
%	2. Raman spectra are interpreted within \textit{ab initio} lattice dynamics calcultion.
	
%	3. There is small but recordable impact of spatial confinement on full widths at half maximum (FWHM) of low-frequency Raman lines of the pSiO$_2$:BNO composite.}

\end{abstract}

\maketitle

%% main text
\section{Introduction}
%\label{}
Crystallization in nanoconfined geometries plays an important role in many natural and technological phenomena ranging from biomineralization and frost heave via protein crystallization and geological mineralization to nanomaterial fabrication with artificial porous scaffold structures \cite{Alba-Simionesco2006, Desarnaud2014, Huber2015, Lindstrom2016,Meldrum2020}. The spatial restriction along with the interaction with the surfaces and geometrical disorder often results in peculiar crystalline textures \cite{Meldrum2020, Wallacher2002}, the absence of phases or completely novel phase behavior, when compared to the unconfined, bulk systems \cite{Christenson2001,Huber2004, Alba-Simionesco2006, Zeng2018, Sentker2018, Meissner2019, Enninful2021}

Experimentally, it is quite challenging to scrutinize crystallization processes in nanoporous media. X-ray and neutron diffraction are arguably the most powerful techniques to study confinement-induced formation of preferred oriented crystals in pore space and to scrutinize confinement effects on the static structure \cite{Sentker2018,Zeng2018, Sentker2019}. Theoretically, it was shown, \textit{e.g.}, that energy band spectrum and consequently electrical transport, thermoelectric and optical properties undergo substantial transformations at decreasing the sample dimensionality from 3D to periodic 1D one \cite{Shi,LHP,Tang,Hosseini,porSi}.

Raman spectroscopy has been widely used for decades to determine vibrational spectra of molecular systems and bulk crystals. Over the last decade, this experimental technique has also been actively used for probing the nanoscale structures such as thin films \cite{Dasi}, nanofiberrs and nanowires \cite{Sharikova,Dhara}, quantum dots \cite{Trajic}, nanoparticles \cite{Krajczewski}. In context of nanocomposite materials, Raman scattering is a very efficient method for structural identification and characterization of nanocrystals confined into nanochannels of host inorganic matrices. Recently we studied silica nanopore-confined KH$_2$PO$_4$ crystals by Raman scattering \cite{Shchur_KDP} and could show that this technique is particularly suitable to explore the static structure (crystal symmetry) in confinement. It was found that the bulk tetragonal symmetry of KH$_2$PO$_4$ is mainly preserved within nanopores approximately 12 nm across at room temperature. Although a certain redistribution of Raman line intensities in nanoconfined KH$_2$PO$_4$ crystals was observed, the overall changes in the Raman spectrum of nanoconfined KH$_2$PO$_4$ are rather small compared to the bulk crystal. Such conserved character of the lattice dynamics should be attributed to a complexity of the phonon spectrum of this tetragonal crystal and hence to the various types of crystal bonding.

By contrast a crystal of higher symmetry, e.g. of cubic symmetry, with simpler type of bonding for which a degeneration of phonon frequencies is inherent, confinement effects could result in the eventual lift of degenerated phonon mode frequencies. Therefore, we selected barium nitrate Ba(NO$_3$)$_2$, hereafter denoted BNO, for this study. It can be embedded in nanoporous media from saturated water solutions. Note that a peculiarity of the BNO Raman spectrum is a very strong line at $\sim$1050 cm$^{-1}$ with a very small line width of 1.2-2.5 cm$^{-1}$. It turned out that the usage of the stimulated Raman scattering (SRS) effect and multipass optical resonator scheme may significantly amplify the Raman Stokes (or anti-Stokes) component at $\sim$1050 \cm to a level sufficient for laser source radiation (Raman shifter) \cite{Karpukhin,Murray,Zverev}. Application of the stimulated Raman scattering for $\lambda$=1.318 $\mu$m pump laser line enables near 50 $\%$ efficiency of Stokes energy conversion into $\lambda$=1.535 $\mu$m line \cite{Murray} and near 26 $\%$ energy conversion efficiency for the transformation of the $\lambda$=0.532 $\mu$m laser radiation to the $\lambda$=0.56 $\mu$m first Stokes component \cite{Eremenko}.

Here, we explore nanconfined BNO crystals. We measure polarized Raman spectra of both pSiO$_2$:BNO composites and X-ray diffraction oriented bulk single BNO crystals. To classify and interpret the experimental Raman data we perform a rigorous group-theory analysis and \textit{ab initio} lattice dynamics calculation of the cubic BNO crystals.

\section{Experimental details}
\begin{figure}[ht]
\begin{center}
 \includegraphics [scale=0.63] {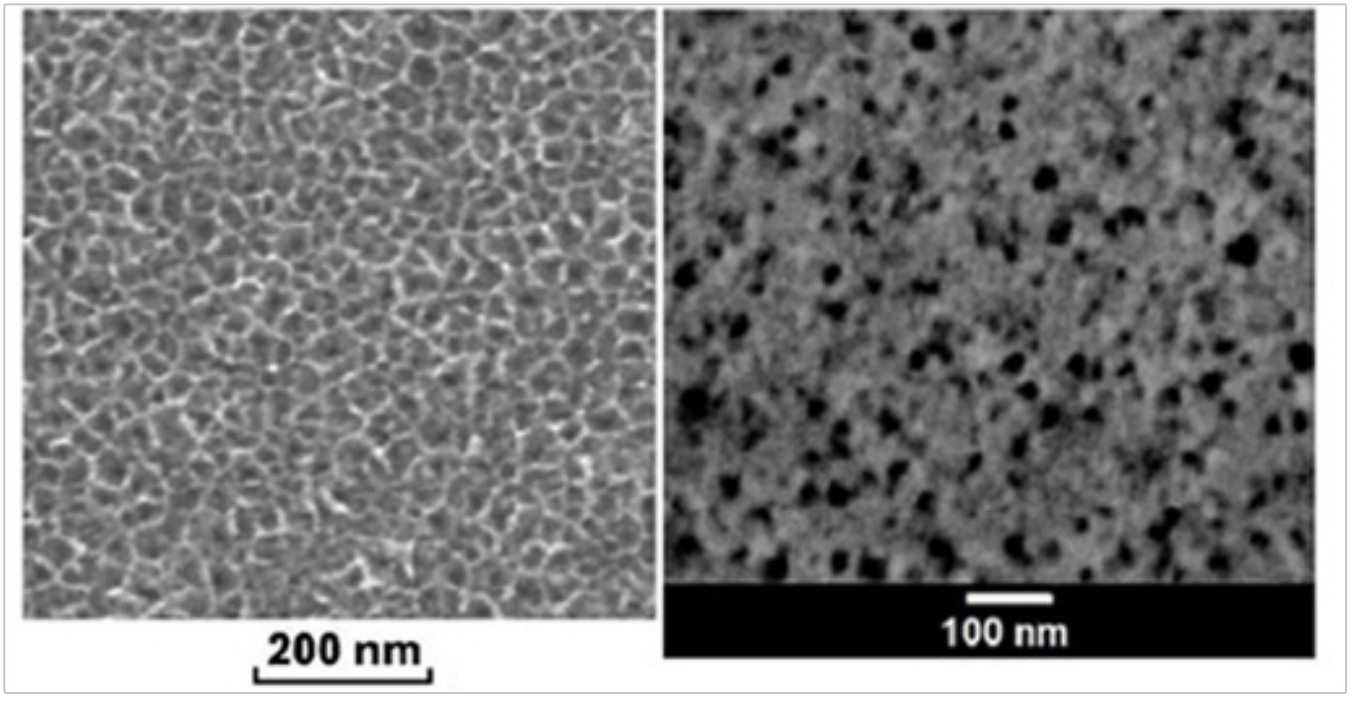}
\end{center}
\caption[]{Scanning electron top-view micrographs of mesoporous pSiO$_2$ membranes with a mean pore diameter of $\sim$12 nm.} \label{sem}
\end{figure}

The mesoporous matrices used for the nanocomposite fabrication represent mesoporous membranes of pSiO$_2$ of thickness L=0.3-0.33 mm with a laterally disordered array of parallel tubular channels of about 12 nm mean diameter, see scanning electron microscopy (SEM) images in Fig.~\ref{sem}. They were synthesized by thermal oxidation of electrochemically etched mesoporous silicon.

BNO nanocrystals were deposited into the nanochannels from saturated water solutions prepared at $\sim$318 K. Subsequent evaporation of water from the pores resulted in crystal formation in pore space.

X-ray diffraction (XRD) patterns were recorded on the diffractometer DRON-3 with a copper anode (K$\alpha$ -line, X-ray wavelength $\lambda$ =1.5406 ${\AA}$).

Confocal Raman microscopy was performed at room temperature in backscattering geometry using a Witec 300 alpha R setup with a spectral resolution of $\sim$2 cm$^{-1}$. The samples were illuminated with 532 nm radiation excited  with a single-mode frequency doubled Nd:YAG laser via a 100 $\mu$m single-mode glass fibre. We used a Zeiss LD EC Epiplan-Neofluar $\times$50/0.55 objective and the laser power at the back of the objective was 12 mW. An edge filter was utilized to separate the Raman signal from the excitation line. Confocality of the Raman signal was achieved via a 50 $\mu$m multi-mode fibre glass between microscope and the Raman spectrometer, where the fibre serves as a pin-hole. The Raman spectrometer was equipped with a holographic grating of 600 lines/mm. As detector a Newton Andor EMCCD camera with 1600$\times$200 pixels was used. Owing to the setup limitations we were able to measure Raman spectra only above 100 cm$^{-1}$.
In the experiments presented here, we typically applied an integration time of $\sim$0.1 sec per spectrum and pixel to improve the signal-to-noise ratio (S/N). To map the surface of pSiO$_2$:BNO composite we used 100$\times$100 pixel Raman scans covering an area of 100$\times$100 $\mu$m. All the data sets were analyzed using Cluster Analysis and Non-Negative Matrix Factorization to demix the measured Raman spectrum in a linear combination of Raman spectra of individual components at each point in a spatial array, namely
\begin{align*}
	S_{T}(x,y)=\sum_{i}w_i(x,y)S_i,
\end{align*}
where \textit{S}$_{T}$(x,y) is the measured Raman spectrum at the point with spatial coordinates (x,y), \textit{S}$_i$ is the Raman spectrum of each component, and \textit{w}$_i$(x,y) is the weighting factor of the \textit{i}-th Raman spectrum at the point (x,y).

\begin{figure}[ht]
	\begin{center}
		\includegraphics [scale=0.35] {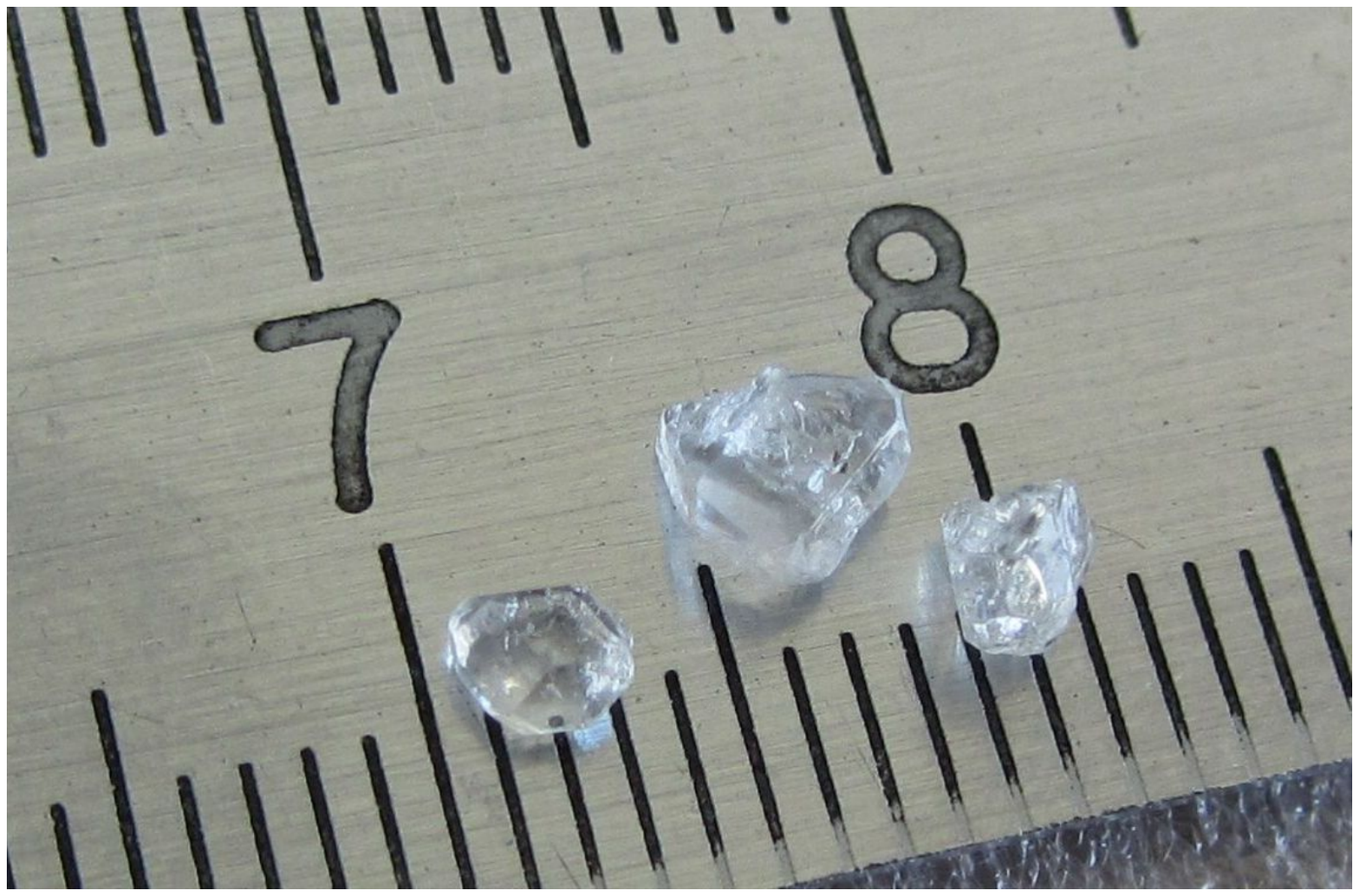}
	\end{center}
	\caption[]{BNO single crystals used for polarized Raman scattering measurements.} \label{BNO_single}
\end{figure}

Polarized Raman spectra were measured from X-ray diffraction oriented BNO single crystals. The BNO samples display a cubic habit with natural well defined isometric (1 0 0) and (1 1 1) planes and small (2 1 0) planes. The samples were several millimeters in size, see Figure~\ref{BNO_single}.

A calculation of the phonon spectrum was performed at the Brillouin zone centre within the density functional perturbation theory (DFPT) using two approaches. Firstly, we used the generalized gradient approximation (GGA) in Perdew-Burke-Ernzerhof parameterization \cite{Perdew}. In the second attempt, we took into consideration the long-range dispersion effects as suggested by Grimme \textit{et al.} (GGA+DFT-D3 method) \cite{Grimme}. All calculations were carried out by the \textit{ab-initio} package ABINIT \cite{Abinit1,Abinit2}. The norm-conserving ONCVPSP pseudopotentials were utilised with Ba(5s$^2$5p$^6$6s$^2$), N(2s$^2$2p$^3$) and O(2s$^2$2p$^4$) valence states. The convergence analysis was performed concerning both the sampling of the Brillouin zone using the Monkhorst-Pack scheme \cite{Monkhorst} and the kinetic energy cut-off for plane-wave calculations. The experimental lattice parameters and atom coordinates resolved at room temperature \cite{Nowotny} were relaxed during structural optimization which was done within the Broyden-Fletcher-Goldfarb-Shanno algorithm \cite{Broyden}. The maximal forces acting on each atom were lower than 1.5$\times$10$^{-8}$ eV/${\AA}$ for both GGA and GGA+DFT-D3 approaches. The 4$\ast$4$\ast$4 grid for Brillouin zone sampling was used and the cut-off energy, E$_{cut}$, was set to 1632 eV with a cut-off smearing of 13.6 eV.

%%%%%%%%%%%%%%%%%%%%%%%%%%%%%%%%%%%%%%%%%
\section{Results and discussion}
\subsection{X-ray diffraction}
%%%%%%%%%%%%%%%%%%%%%%%%%%%%%%%%%%%%%%%%%%%%%%%%%%%%%%%%%%%%%%%%%%%%%%%%%%%%%%%
\begin{figure}[!ht]
\begin{center}
 \includegraphics [scale=0.75] {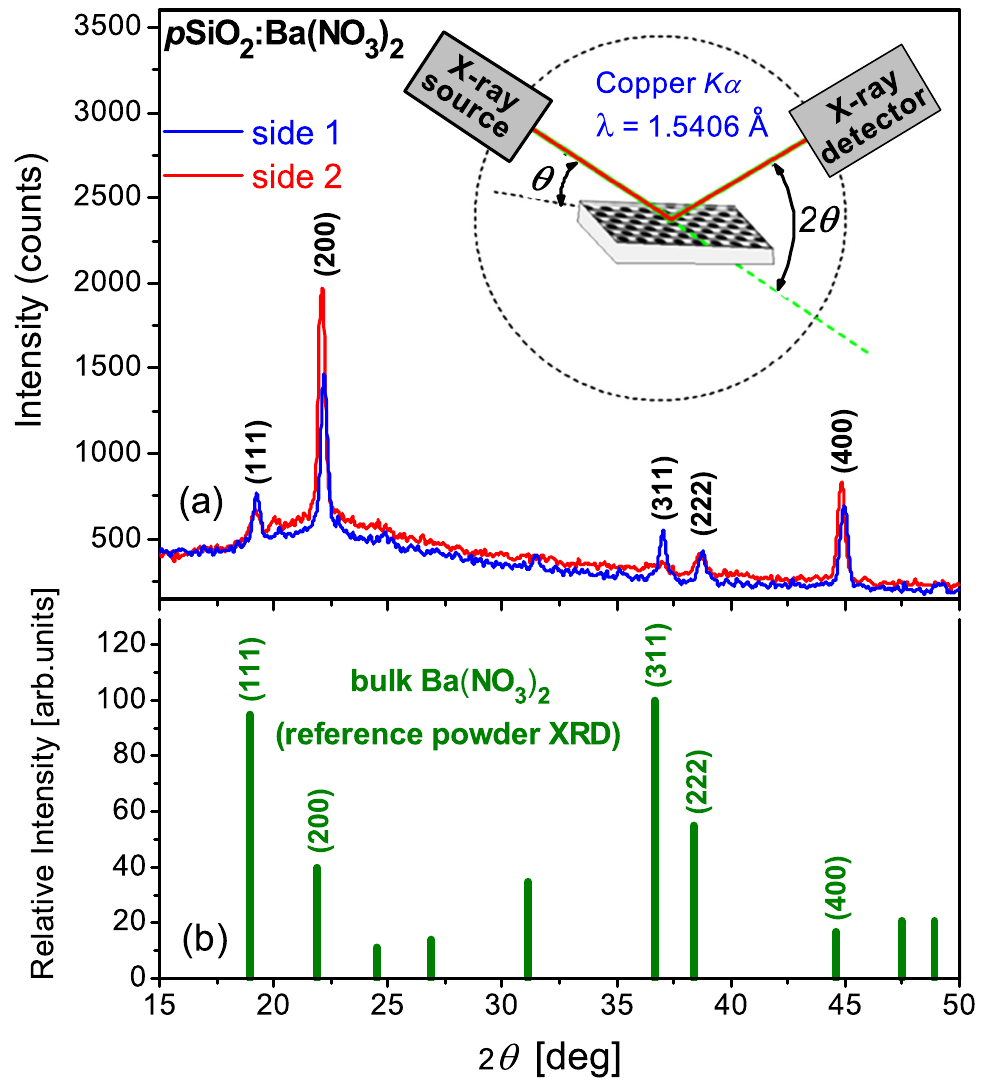}
\end{center}
\caption[]{(a) X-ray diffraction patterns ($\theta$/2$\theta$-scan) recorded from opposite sides of pSiO$_2$:BNO membranes (b) reference bulk X-ray diffraction powder pattern of BNO.} \label{xrd}
\end{figure}
%%%%%%%%%%%%%%%%%%%%%%%%%%%%%%%%%%%%%%%%%%%%%%%%%%%%
In Figure \ref{xrd} the $\theta$/2$\theta$ reflection XRD geometry is presented along with XRD patterns recorded on both sides of the nanocomposite membrane and a bulk reference pattern. The relative Bragg intensities indicate a strong texture with a dominating (100)-orientation ((200)-reflection in XRD pattern) of BNO nanocrystals along the long channel axis. A quantitative comparison of the XRD peak heights recorded for both sides of pSiO$_2$:BNO and the reference powder XRD pattern (see Figure \ref{xrd}(b)), respectively, indicates that a fraction of about 87$\%$ of BNO nanoncrystals is preferentially in (100)-orientation along the long channel axis. For comparison, the nanocrystals of two other crystallographic orientations present in the XRD patterns, i.e. (111) and (311), enter into the texture with a fraction of about 8$\%$ and 5$\%$, respectively.

By extracting the full widths at half maximum (FWHM) of XRD peak (200) (Figure \ref{xrd}) and applying the
Scherrer equation \cite{Patterson} one obtains the size of embedded BNO crystallites  of about 30 nm along nanochannels, i.e. perpendicular to the silica membrane faces. This value, on the other hand, contrasts with their micrometric lateral size, which is about the lengths of the dendrites spreading between the different channels observed, particularly, by means of polarized and Raman microscopy.

Note that the XRD patterns of the distinct membrane sides differ somewhat, which indicates that the BNO texture is inhomogeneous along the nanochannels. This could be related to a conical shape of the nanochannels that is known to occur during the electrochemical etching process of the mesoporous silicon matrix \cite{Waszkowska2021SHG}.

\subsection{Lattice dynamics of BNO: symmetry considerations}
\begin{figure}[ht]
\begin{center}
 \includegraphics [scale=0.80] {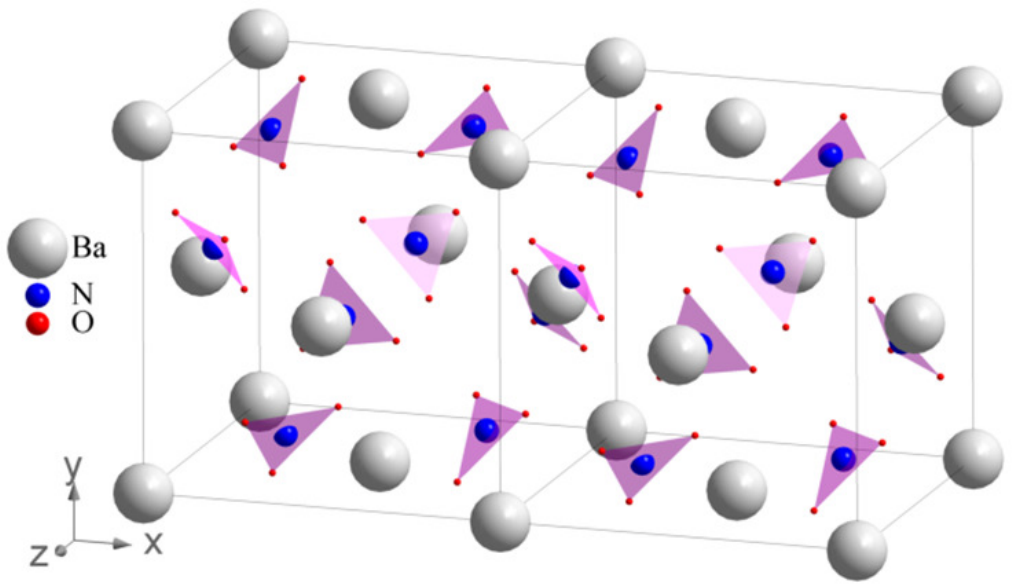}
\end{center}
\caption[]{Crystal structure of BNO at room temperature \cite{Nowotny} (space group \textit{Pa$\bar{3}$ }, No. 205).} \label{BNO_str}
\end{figure}

BNO crystallizes in the cubic \textit{Pa$\bar{3}$ }(No. 205) space group. There are four structural units (Z=4) in the unit cell (see Fig.~\ref{BNO_str}). The main structural characteristic feature of this crystal is planar covalently bonded NO$_3$ triangles linked with Ba ions via ionic electrostatic forces. 108 normal vibrational modes are classified according to the irreducible representations of the \textit{Pa$\bar{3}$} group in Brillouin zone center as follows,

$\Gamma$(108)= 4A$_g$(R)+4E$_g$(R)+12T$_g$(R)+5A$_u$+5E$_u$+15T$_u$(IR)

R and IR imply the Raman or IR activity of the corresponding irreducible representations (irreps). The modes of both A$_u$ and E$_u$ symmetry should not be visible in Raman and IR spectra. Note that the pairs E$_g$, E$_u$ and T$_g$, T$_u$ are two- and three-dimensional irreps, therefore all modes transformed according to these irreps are two- and three-fold degenerated ones.

Four Ba atoms create 12 translational modes. Eight covalently bonded planar NO$_3$ molecular groups are involved in 96 normal modes, namely:
24 translations, 24 librations (rotations) and 48 internal N-O vibrations. Using the atomic coordinates, corresponding Wickoff positions \cite{Nowotny} and group-theory methods \cite{Poulet,Nakamoto} one may find the contribution of every kind of atoms to the total normal vibrations classified according to the following irreps

Ba(4a): A$_u$+E$_u$+3T$_u$

N(8c): A$_g$+E$_g$+3T$_g$+A$_u$+E$_u$+3T$_u$

O(24d): 3A$_g$+3E$_g$+9T$_g$+3A$_u$+3E$_u$+9T$_u$.

\begin{figure}[ht]
\begin{center}
 \includegraphics [scale=0.76] {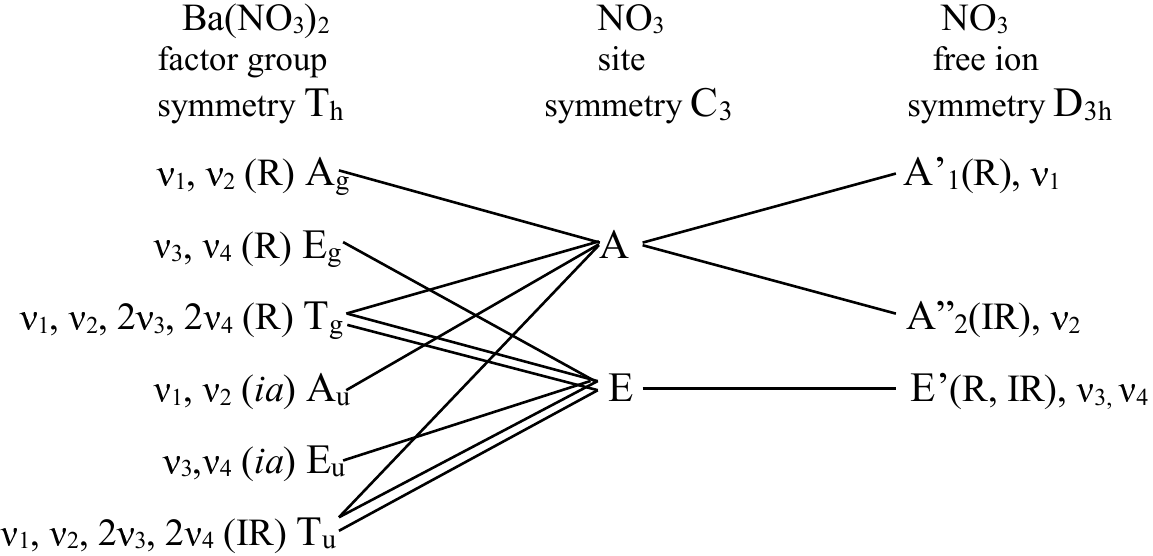}
\end{center}
\caption[]{Correlation diagram for internal modes of the NO$_3$ group ($\nu_1$, $\nu_2$, $\nu_3$, $\nu_4$) in a BNO crystal. The abbreviation \textit{ia} implies that the modes of A$_u$ and E$_u$ species are inactive both in Raman and IR spectra.} \label{cor_diag}
\end{figure}

All internal vibrations of the free NO$_3$ covalently bonded groups are very well known. According to Nakamoto's textbook \cite{Nakamoto}, they are as follows, $\nu_1$= 1068 cm$^{-1}$ (R), $\nu_2$= 832 cm$^{-1}$ (IR), $\nu_3$= 1385-1405 cm$^{-1}$ (R, IR), $\nu_4$= 692-724 cm$^{-1}$ (R, IR). Applying the group-theory methods one may construct the correlation diagram for internal modes of the NO$_3$ group ($\nu_1$, $\nu_2$, $\nu_3$, $\nu_4$) which manifests the relation between the factor group of the BNO crystal and the free NO$_3$ ion symmetry (see Figure \ref{cor_diag}). Using this figure, one may easily predict the number and type of internal modes transformed according to each irrep.

Summarizing the available group-theory information we come to the final symmetry classification of phonon modes of BNO crystals (see Table \ref{Classif}). The frequencies of all phonon modes may be approximately distributed over four ranges. The lowest frequency range, less than 200 cm$^{-1}$, corresponds to external vibrations (acoustic, translations and librations). The region of 700 to 800 cm$^{-1}$ comprises internal $\nu_2$, $\nu_4$ vibrations of NO$_3$ groups, whereas the two high frequency ranges near 1050 and 1380 cm$^{-1}$ correspond to $\nu_1$ and $\nu_3$ internal NO$_3$ modes, respectively. In Table \ref{Classif} we also indicate the total number of Raman active modes that may be expected for observation in polarized Raman spectra in each peculiar range.

%%%%%%%%%%%%%%%%%%%%%%%%%%%%%%%%%%%%%%%%%%%%%%%%%%
\begin{table*}
\caption {Normal mode classification of BNO, sp. gr. \textit{Pa$\bar{3}$}. The components of Raman polarizability tensor are indicated near Raman active irreps.}
 \label{Classif}
\footnotesize
%\hspace*{-2em}
%\small
\begin{tabular}{lcccc}
\\\hline
Irreps&$<$200 cm$^{-1}$&700-800 cm$^{-1}$&$\sim$1060 cm$^{-1}$&$>$1360 cm$^{-1}$ \\
\hline
4A$_g$(X$^2$+Y$^2$+Z$^2$)&2 (1 transl.+1 libr.)&$\nu_2$&$\nu_1$&\\
4E$_g$(2Z$^2$-X$^2$-Y$^2$, X$^2$-Y$^2$)&2 (1 transl.+1 libr.)&$\nu_4$&&$\nu_3$\\
12T$_g$(XY, YZ, XZ)&6 (3 transl.+3 libr.)&$\nu_2$, 2$\nu_4$&$\nu_1$&2$\nu_3$\\
5A$_u$&3 (2 transl.+1 libr.)&$\nu_2$&$\nu_1$&\\
5E$_u$&3 (2 transl.+1 libr.)&$\nu_4$&&$\nu_3$\\
15T$_u$&9 (1 acoust. +5 transl.+&$\nu_2$, 2$\nu_4$&$\nu_1$&2$\nu_3$\\
&3 libr.)&&&\\
\hline
$\sum$ Raman&10&5&2&3\\
active modes&&&&\\
A$_g$+E$_g$&4&2&1&1\\
T$_g$&6&3&1&2\\
\hline
\end{tabular}
\end{table*}
%%%%%%%%%%%%%%%%%%%%%%%%%%%%%%%%%%%%%%%%%

\subsection{Raman scattering of a BNO single crystal}
How can we use the Table \ref{Classif} for interpreting the Raman spectra of an oriented single crystal recorded in back-ward scattering geometry? According to the symmetry of the Raman polarizability tensor (see Table \ref{Classif}) all geometries of Raman scattering experiments with a parallel orientation of both incident and scattered beam, namely Y(XX)-Y, Z(XX)-Z, X(YY)-X, Z(YY)-Z, X(ZZ)-X, Y(ZZ)-Y correspond simultaneously to A$_g$ and E$_g$ irreps. All experimental geometries with perpendicular polarization of incident and scattered beam, namely Y(XZ)-Y, Z(XY)-Z, X(YZ)-X  correspond to T$_g$ irrep. Note that in a cubic crystal the X, Y, Z notation of crystallographic axes is provisional since all these axes are identical. In other words, the use of a standard polarized Raman scattering technique cannot discern the phonons of A$_g$ and E$_g$ species in the cubic crystal of \textit{Pa$\bar{3}$ }symmetry. However, the application of some experimental trick suggested by Lockwood \cite{Lockwood} with a rotation of the incident beam polarization by 45$\degree$ around Z axis to X' direction (angle between X and X' axes is 45$\degree$, X'$\|$[110]) and observation of scattered light polarized along Y' axis orthogonal to X' one (Y'$\|$[1-10]) allows the transformation of the Raman polarizability tensor from the form  written in X, Y, Z coordinate system
\begin{equation*}
A_{g}:
\begin{pmatrix}
 a& &  \\
  & a& \\
  &  & a \
\end{pmatrix},
E_{g} =
\begin{pmatrix}
 b& &  \\
  & b& \\
  &  &-2b \
\end{pmatrix},
\begin{pmatrix}
 -\sqrt{3}b& &  \\
  & \sqrt{3}b& \\
  &  & \
\end{pmatrix},
\end{equation*}

\begin{equation*}
T_{g}:
\begin{pmatrix}
 & &  \\
  & &d \\
  & d& \
\end{pmatrix},
\begin{pmatrix}
 & &d  \\
  & & \\
d &  & \
\end{pmatrix},
\begin{pmatrix}
 &d &  \\
 d & & \\
  &  & \
\end{pmatrix}
\end{equation*}
to another form set in X', Y', Z coordinate system:
\begin{equation*}
A_{g}:
\begin{pmatrix}
 a& &  \\
  & a& \\
  &  & a \
\end{pmatrix},
E_{g} =
\begin{pmatrix}
 b& &  \\
  & b& \\
  &  &-2b \
\end{pmatrix},
\begin{pmatrix}
 &\sqrt{3}b &  \\
\sqrt{3}b & & \\
  &  & \
\end{pmatrix},
\end{equation*}

\begin{equation*}
T_{g}:
\begin{pmatrix}
 & &\frac{d}{\sqrt{2}} \\
  & &\frac{d}{\sqrt{2}} \\
 \frac{d}{\sqrt{2}}&\frac{d}{\sqrt{2}}& \
\end{pmatrix},
\begin{pmatrix}
 & &\frac{d}{\sqrt{2}}  \\
  & &-\frac{d}{\sqrt{2}} \\
\frac{d}{\sqrt{2}}&-\frac{d}{\sqrt{2}} & \
\end{pmatrix},
\begin{pmatrix}
d& &  \\
 &-d & \\
  &  & \
\end{pmatrix}.
\end{equation*}
Therefore the usage of any parallel back-ward Raman scattering geometry, e.g. Z(XX)-Z or Z(YY)-Z, and diagonal Z(X'Y')-Z geometry with the subsequent subtraction of two experimental Raman spectra allows a separation of A$_g$ and E$_g$ constituents.

\begin{figure*}[!ht]
\begin{center}
\includegraphics[scale=0.3, angle=0]{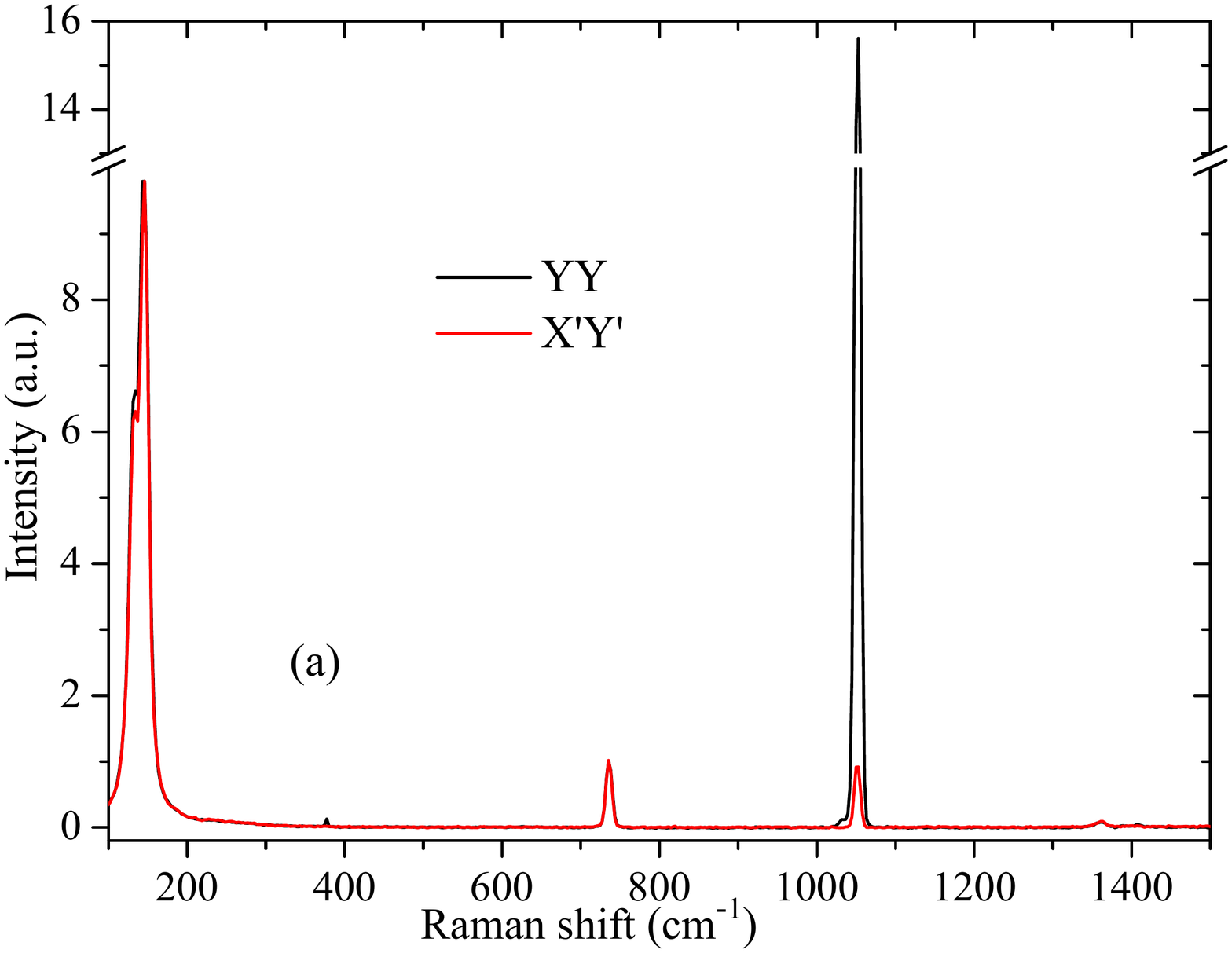}~~ \includegraphics[scale=0.3, angle=0]{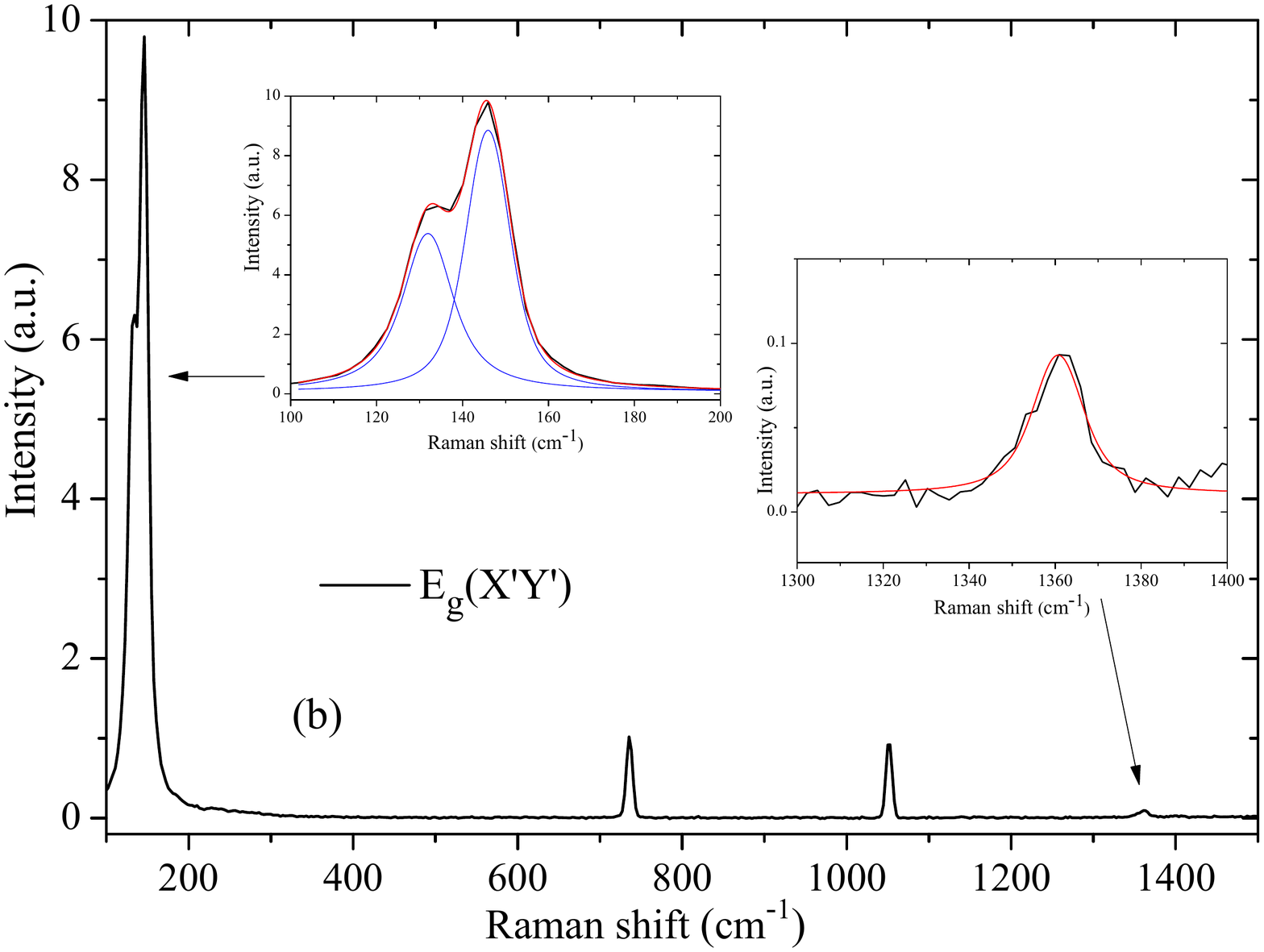}\\
\includegraphics[scale=0.3, angle=0]{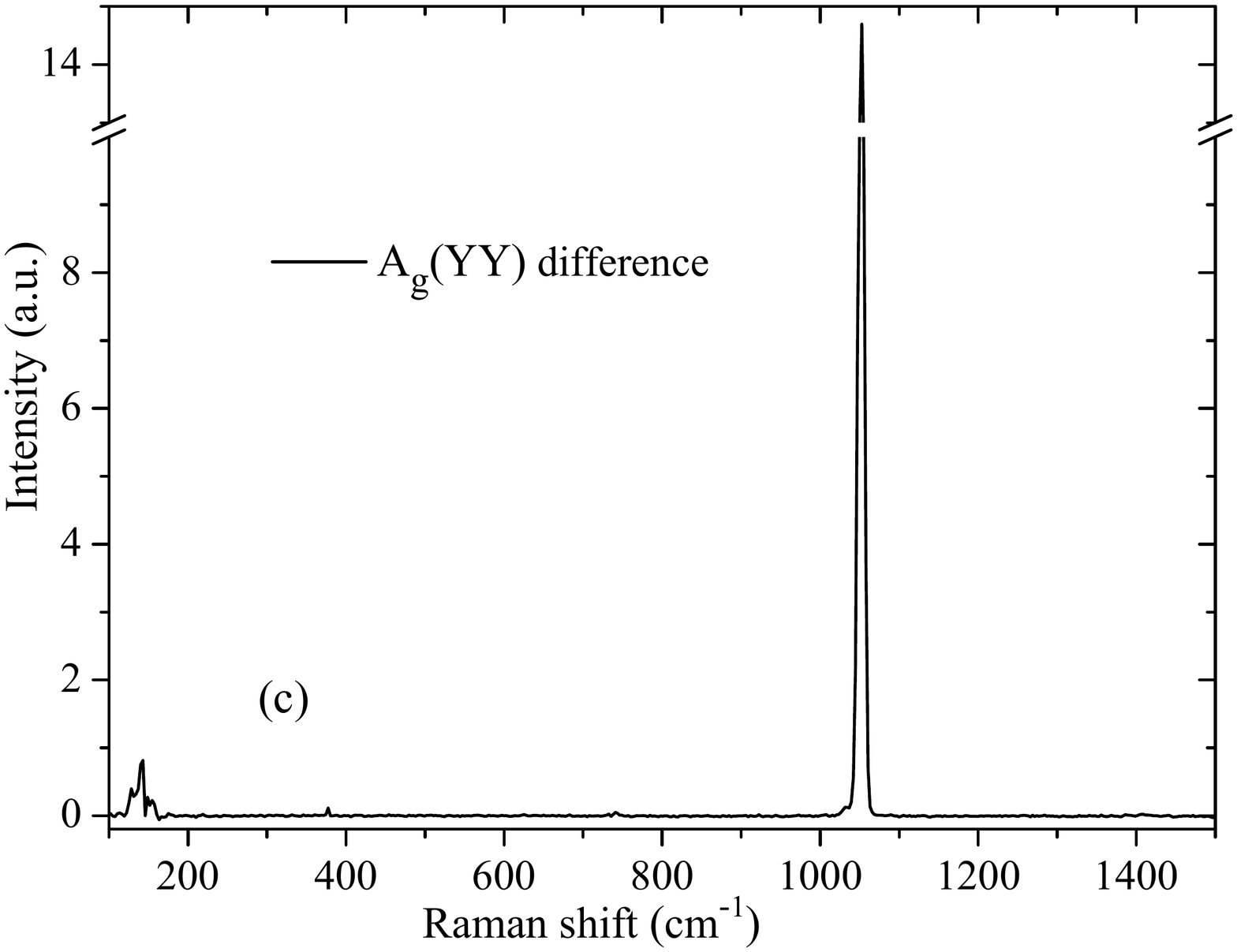}~~ \includegraphics[scale=0.3, angle=0]{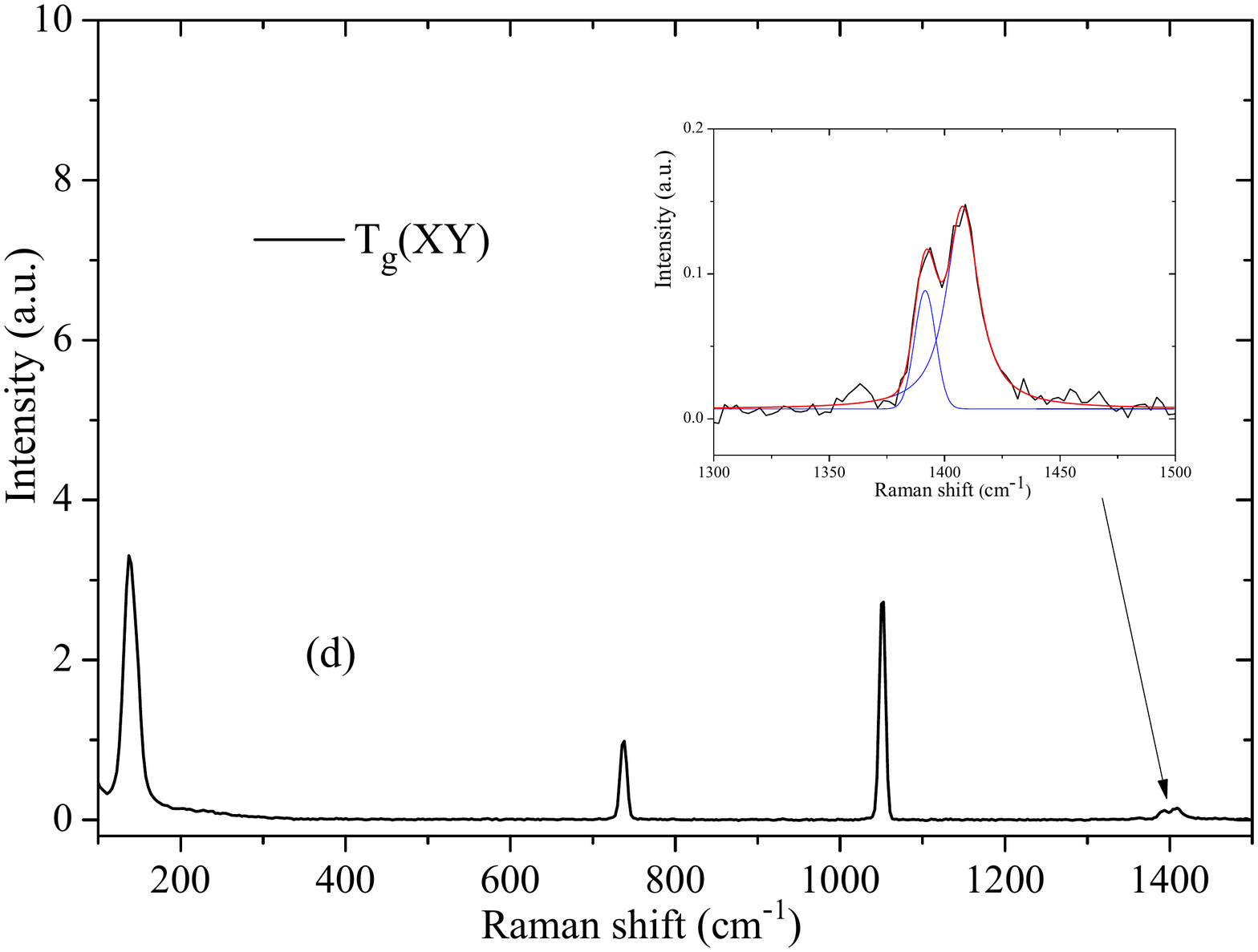}\\
\end{center}
\caption[]{Polarized Raman spectra of a BNO single crystal. (a) The Z(YY)-Z (A$_g$+E$_g$, black curve) and Z(X'Y')-Z (E$_g$, red curve) Raman spectra; (b) the E$_g$ spectrum recorded in rotated Z(X'Y')-Z geometry. Insets show some parts of this spectrum with line shape fitted assuming a Lorentzian profile of constituent Raman lines (blue curves); (c) the A$_g$ spectrum obtained as the result of simple subtraction between Z(YY)-Z and Z(X'Y')-Z spectra depicted in Figure \ref{Ram}(a); (d) the Z(XY)-Z (T$_g$) spectrum, inset shows low intensity high frequency part with two Lorentzian constituents.} \label{Ram}
\end{figure*}

Figure \ref{Ram}(a) presents Raman spectra of a BNO crystal recorded at room temperature in Z(YY)-Z geometry. This spectrum contains the phonon modes of both A$_g$ and E$_g$ symmetries. The E$_g$ spectrum obtained in the rotated Z(X'Y')-Z geometry is depicted in Figure \ref{Ram}(b). In the same figure we present the low- and high-frequency constituents revealed after the line shape fitting using the Lorentzian profile. The result of subsequent subtraction of Z(YY)-Z and Z(X'Y')-Z spectra yields the totally symmetric A$_g$ spectrum  (see Figure \ref{Ram}(c)). The modes of T$_g$ symmetry become visible in Z(XY)-Z geometry (Figure \ref{Ram}(d)). The eye-catching feature of these spectra is their comparative poorness. This especially concerns to A$_g$ and T$_g$ spectra, Figures \ref{Ram}(c, d), in which, instead of four and twelve theoretically allowed modes, we experimentally detected one and six lines, respectively. Note that a few low intensity modes visible below 200 cm$^{-1}$ in A$_g$ spectrum (Figure \ref{Ram}(c)) may be the result of a leakage effect
which becomes apparent after subtraction of Z(YY)-Z and Z(X'Y')-Z spectra. The single narrow high-intensity A$_g$ mode at 1052 cm$^{-1}$ is exactly the line used in SRS lasers as the radiation source. According to Table \ref{Classif} this internal stretching $\nu_1$ vibration of the NO$_3$ group should manifest itself as Raman activity in both A$_g$ and T$_g$ spectra and, in fact, it is clearly observable in both spectra. However, the line intensity of the corresponding T$_g$ $\nu_1$ line is almost five times smaller compared with its A$_g$ counterpart (see Figure \ref{Ram}(c, d)). All four theoretically predicted modes of E$_g$ symmetry are clearly discernible in our Z(X'Y')-Z spectrum (Figure \ref{Ram}(b)). Note that according to Table \ref{Classif} the highest frequency NO$_3$ internal modes $\nu_3$ should be observable in E$_g$ and T$_g$ spectra above 1360 cm$^{-1}$. Our experimental observation nicely corroborates this general symmetry demand. As seen in Figures \ref{Ram}(b,d) one and two low-intensity modes are detected near 1400 cm$^{-1}$ in E$_g$ and T$_g$ geometries, respectively. The medium intensity mode detected near $\sim$1053 cm$^{-1}$ in Z(X'Y')-Z spectrum (Figure \ref{Ram}(b)), most probably, is not the intrinsic E$_g$ mode but should be treated as the leakage of the very strong mode of A$_g$ symmetry. To prove this observation and to achieve a rigorous understanding of all vibrational modes we have undertaken a first-principles analysis of BNO lattice dynamics.

\subsection{Lattice dynamics: \textit{Ab initio} calculation}
The comparison between phonon frequencies calculated in the Brillouin zone center within two approaches, GGA and GGA+DFT-D3, and experimental ones detected by us in polarized Raman scattering on the bulk BNO single crystal are presented in Table \ref{Compar}. In this table the experimental frequency of A$_g$ symmetry corresponds to that obtained after subtraction of two Raman spectra, Z(YY)-Z and Z(X'Y')-Z. IR active phonon modes transformed according to T$_u$ irrep are compared with previously published IR data \cite{Bon}. For all T$_u$ symmetry modes we indicate the frequencies of both transverse TO and longitudinal LO types of polar optical vibrations. As seen in this table, the long-range dispersion corrections used in GGA+DFT-D3 method \cite{Grimme} change little the phonon frequencies calculated in the Brillouin zone center revealing practically the same total deviation between calculated and experimentally detected frequencies, i.e., 3.3 $\%$ for GGA+DFT-D3 against 3.2 $\%$ for GGA method. Based on the analysis of calculated \textit{eigen}vectors we performed the mode assignment at the $\Gamma$ point. It turned out that the classification of calculated modes perfectly agrees with general symmetry requirements presented in Table \ref{Classif}. This concerns both the number of external and internal modes redistributed over
%%%%%%%%%%%%%%%%%%%%%%%%%%%%%%%%%%%%%%%%%%%%%%%%%%%%%%%%%%%%%%%%%%%%%%%%%%%%%%%%%%%
\begin{table*}
\caption{\small Comparison between the phonon frequencies calculated at the $\Gamma$ point (sp. gr. Pa$\bar{3}$) and the experimental Raman frequencies detected in the bulk BNO single crystal. RS corresponds to Raman scattering. D3 refers to GGA+DFT-D3 method. All frequencies are indicated in cm$^{-1}$. $\nu_1$, $\nu_2$, $\nu_3$, $\nu_4$ refer the NO$_3$ internal modes. All modes of frequencies below $\sim$184 cm$^{-1}$ are external lattice vibrations. TO and LO correspond to transverse and longitudinal optic modes, respectively.}
 \label{Compar}
%\hspace*{-3em}
%\small
\footnotesize
\begin{tabular} {ccccccccrcrccc}
\\ \hline
\multicolumn{3}{c}{4A$_g$} &5A$_u$& \multicolumn{3}{c}{4E$_g$}&5E$_u$& \multicolumn{3}{c}{12T$_g$}&
\multicolumn{3}{c}{15T$_u$(TO-LO)}\\ \hline
\multicolumn{2}{c}{calcul.}&RS  & calcul.&\multicolumn{2}{c}{calcul.}&RS &calcul. &\multicolumn{2}{c}{calcul.}&RS&\multicolumn{2}{c}{calcul.}&IR\\
GGA&D3&&D3&GGA&D3&&D3&GGA&D3&&GGA&D3&Ref.\cite{Bon}
\\ \hline
48.9&41.8&&80.9&111.6&119.4&132&48.2&85.3&82.3&&acoust.&acoust.&\\
169.9&183.5&&116.0&145.6&146.3&146&115.3&94.9&96.6&&50.9-51.0&52.2-52.3&\\
$\nu_2$ 792.3&791.8&&165.0&$\nu_4$ 711.4&712.6&736&138.4&129.7&131.9&&81.3-85.3&78.5-80.9&82-90\\
$\nu_1$ 1037.2&1037.6&1052&$\nu_2$ 789.2&$\nu_3$ 1328.7&1329.8&1361&$\nu_4$ 710.7&140.3&143.4&137&93.3-94.6&94.0-95.5&92-94\\
&&&$\nu_1$ 1036.9&&&&$\nu_3$ 1365.5&150.8&157.4&147&106.4-106.5&113.0-113.1&98-101\\
&&&&&&&&165.2&175.9&&126.4-129.7&134.3-138.4&\\
&&&&&&&&$\nu_4$ 712.5&713.9&&134.7-140.3&140.2-143.4&\\
&&&&&&&&$\nu_4$ 714.9&716.4&737&155.2-163.2&165.7-168.1&150-174\\
&&&&&&&&$\nu_2$ 791.3&790.9&&166.5-169.9&169.8-175.9&180-192\\
&&&&&&&&$\nu_1$ 1036.3&1036.4&1051&$\nu_4$ 709.9-710.1&711.1-711.2&\\
&&&&&&&&$\nu_3$ 1356.5&1358.4&1392&$\nu_4$ 710.9-711.4&712.0-712.6&731-734\\
&&&&&&&&$\nu_3$ 1374.9&1376.1&1407&$\nu_2$ 790.1-790.8&789.5-790.2&818-822\\
&&&&&&&&&&&$\nu_1$ 1036.4-1036.4&1036.6-1036.6&\\
&&&&&&&&&&&$\nu_3$ 1319.0-1328.7&1320.1-1329.8&1349-1358\\
&&&&&&&&&&&$\nu_3$ 1375.5-1394.9&1377.8-1397.4&1426-1432\\
\hline
\end{tabular}
\end{table*}
%%%%%%%%%%%%%%%%%%%%%%%%%%%%%%%%%%%%
irreps and the precise type of internal vibrations, $\nu_1$, $\nu_2$, $\nu_3$, $\nu_4$, transformed according to peculiar irrep. There is practically no mixing of different type modes, in contrast to the case of recently investigated nanocrystals of KH$_2$PO$_4$ \cite{Shchur_KDP} and other representatives of KH$_2$PO$_4$ type crystals \cite{CDP2,DRDP,TDP,LHP2}.  All phonon modes with frequencies lower than 170 cm$^{-1}$ are external lattice vibrations. The bending $\nu_4$ and $\nu_2$ vibrations of NO$_3$ planar groups are located near 712 and 790 cm$^{-1}$, respectively, whereas the stretching $\nu_1$ and $\nu_3$ modes have frequencies near 1037 and 1328-1377 cm$^{-1}$, respectively. As follows from the calculated phonon frequencies (Table \ref{Compar}) no phonon mode of E$_g$ symmetry is expected between 711 and 1329 cm$^{-1}$. This means the medium intensity line visible near $\sim$1053 cm$^{-1}$ in Z(X'Y')-Z spectrum (Figure \ref{Ram}(b)) should be treated as the leakage of the very strong totally symmetric mode of A$_g$ symmetry, as was supposed in previous section based on the general symmetry consideration.

It is worth noting that the internal NO$_3$ mode frequencies observed in our Raman experiment appeared to be somewhat smaller than the corresponding frequencies of $\nu_1$, $\nu_2$, $\nu_3$, $\nu_4$ modes indicated in Nakamoto's textbook \cite{Nakamoto}. A similar phenomenon was observed for internal modes of planar BO$_3$ groups in the Raman spectrum of Bi$_3$TeBO$_9$ crystal \cite{BTBO}. The calculated frequencies of most of the internal NO$_3$ modes are also underestimated regarding the experimentally detected ones. This conclusion becomes obvious after careful examination of Table \ref{Compar}. Another peculiar feature of our calculation is the rather small values of the calculated LO-TO splitting of the T$_u$ mode frequencies as compared with those detected in IR reflection spectra \cite{Bon}. Both calculation methods, GGA and GGA+DFT-D3, give much smaller values of LO-TO splitting. This especially concerns the lattice modes near 106(113), 711 and 790 cm$^{-1}$ for which the calculated LO-TO splitting is of one order of magnitude smaller than their experimental counterparts (see Table \ref{Compar}). Since the physical phenomenon of LO-TO splitting is directly related to macroscopic long range electric fields created by charged ions displaced in long-wave LO phonons, one may generally conclude that the current \textit{ab initio} treatment within both GGA and GGA+DFT-D3 methods underestimates the long range dipole-dipole interactions.

\subsection{Polarization microscopy and Raman scattering of nanoconfined BNO}

\begin{figure}[ht]
\begin{center}
 \includegraphics [scale=0.6] {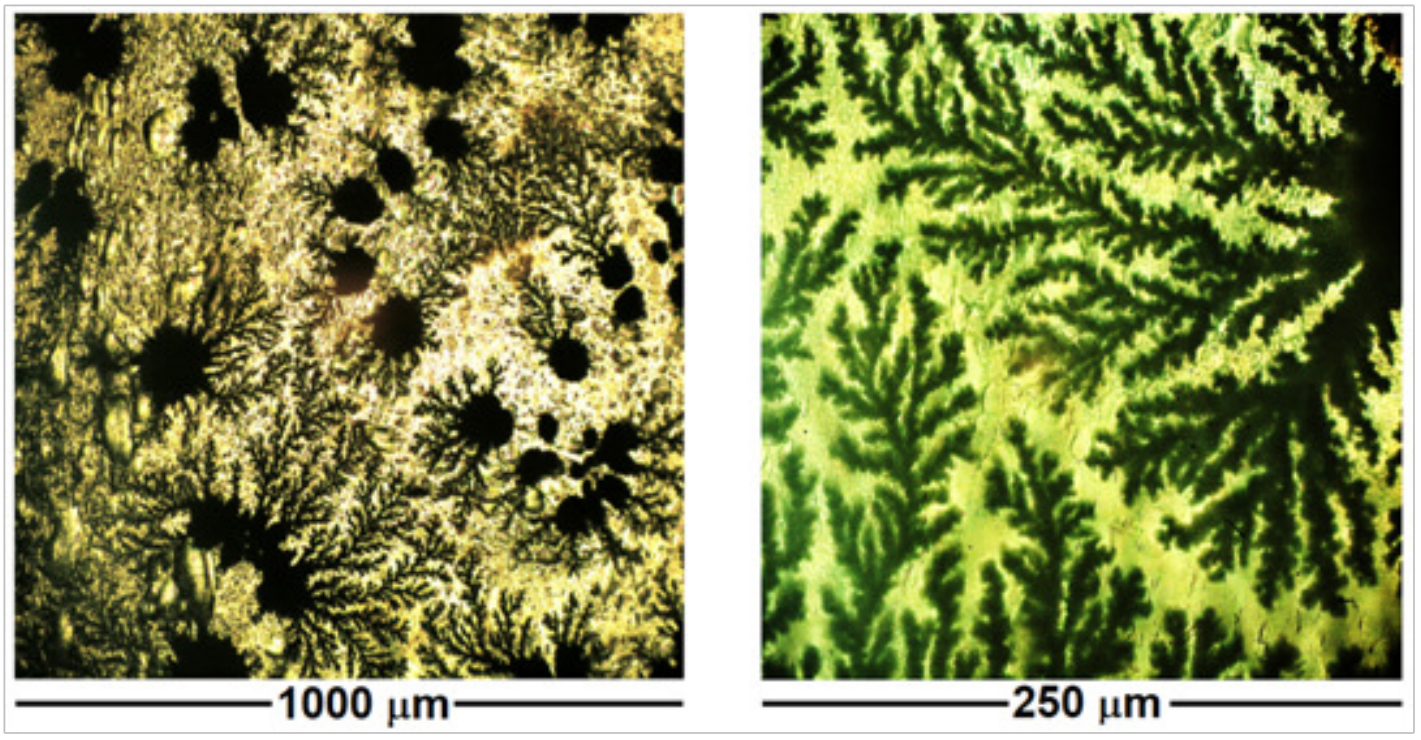}
\end{center}
\caption[]{Polarization microscopy images of pSiO$_2$:BNO at different magnifications.} \label{microsc}
\end{figure}

Surprisingly, the formation of nanocrystals is quite inhomogeneous on macroscopic scales, as evidenced in  polarization microscopy, see images in Figure~ \ref{microsc}. The crystallization in the nanochannels results in micron-scale segregated regions consisting of clustered nanocrystals with identical orientation. These local formations appear deep in the pSiO$_2$ matrix. They range in size from a few tens to about one hundred of microns and are characterized by dendritic-like structure. These crystal growth patterns hint towards propagation of the crystallization front between the tubular nanopores via microporosity in the silica pore walls. This finding is reminiscent of the propagation of cavitation events and of diffusion measurements \cite{Kondrashova2017} in mesoporous silicon that indicate a sizeable porosity of the pore walls. The oxidized version employed here as confining matrix obviously is therefore also rather properly described as an anisotropic 3-D porous medium than as a set of independent tubular nanopores.

\begin{figure*}[ht]
\begin{center}
\includegraphics[scale=0.15]{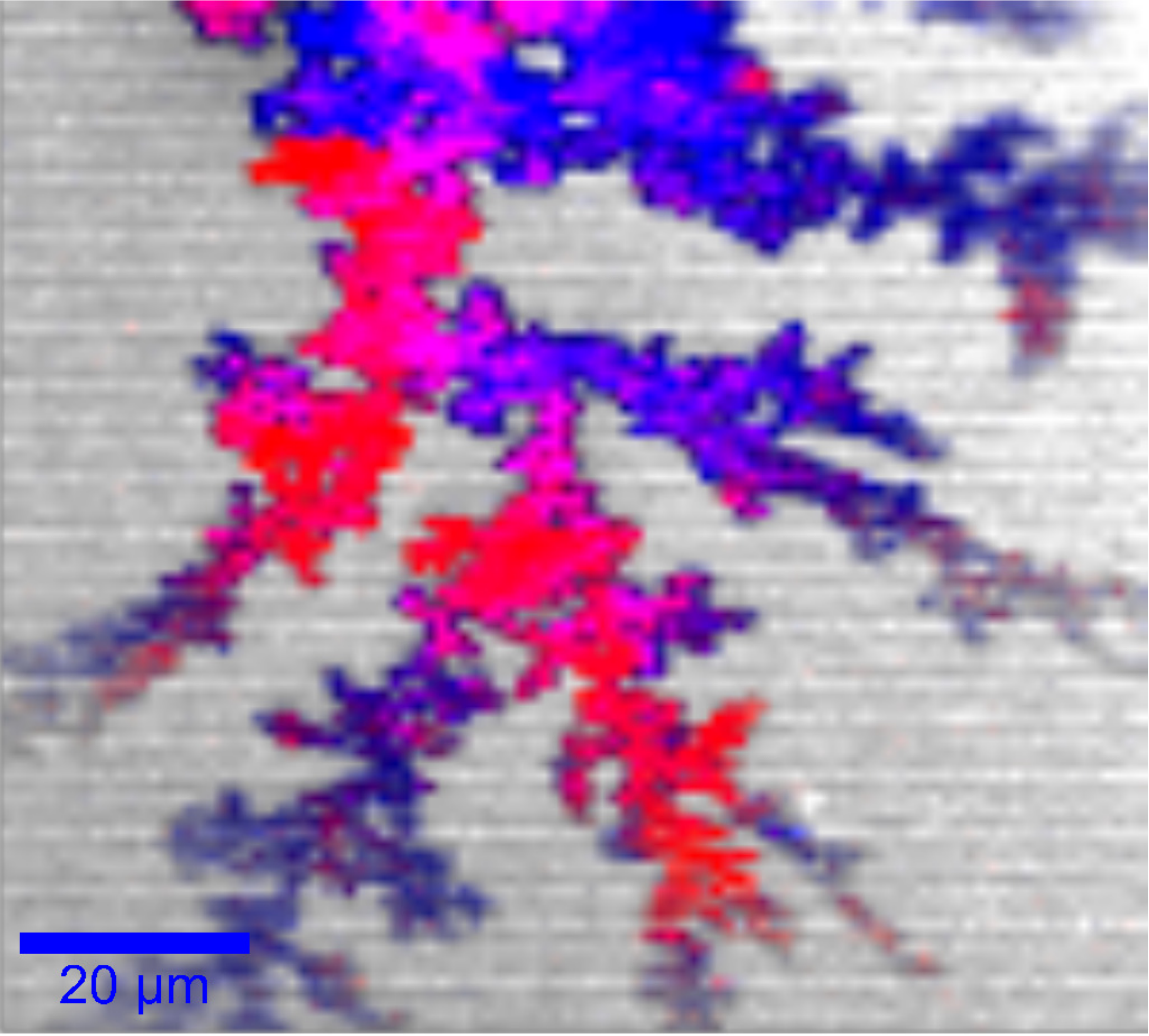}~~ \includegraphics[scale=0.15]{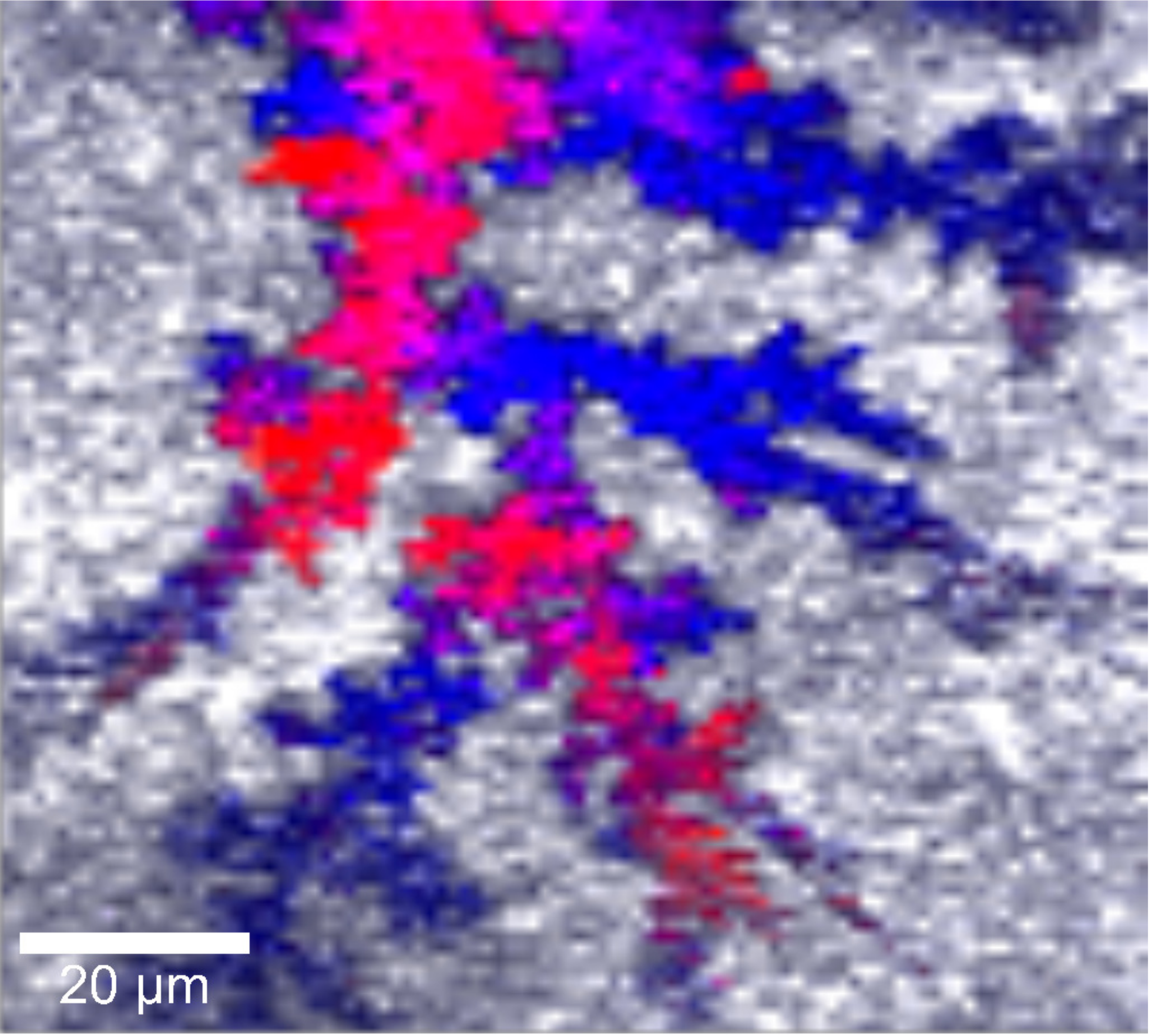}\\
\end{center}
\caption[]{Raman scattering mapping of the surface of pSiO$_2$:BNO in polarized light for parallel (left) and perpendicular (right) polarizations of incident and reflected beams.} \label{Map}
\end{figure*}

We have carried out a careful Raman scattering mapping of pSiO$_2$:BNO sample at different points of its surface with polarized light. Typical patterns of recorded Raman scattering map obtained at parallel (00) and perpendicular (090) polarizations of incident and reflected beams in backscattering geometry are presented in Fig.~\ref{Map}.
%%%%%%%%%%%%%%%%%%%%%%%%%%%%%%%%%%%%%%%%%%%%%%%%%%%%%%%%%%%%%%%%%%%%%%%%%%%%%%%%%%%
There is an impressive similarity between Figures \ref{microsc} and \ref{Map}. Raman scattering mapping well reproduces the dendrite-like structure of the newly synthesized compound embedded into pSiO$_2$ host matrix. A second prominent feature of our observation is the large homogeneous regions of this dendrite-type structure with the uniform Raman spectra which are depicted in Figure \ref{Map}. This figure shows the weighting factors for the two demixed spectra of BNO (one in blue and the other in red) and the one corresponding to the SiO$_2$ matrix (gray). The magenta pixels are present when the weighting factor of the two BNO components, red and blue, are similar.

\begin{figure*}[ht]
\begin{center}
\includegraphics[scale=0.27]{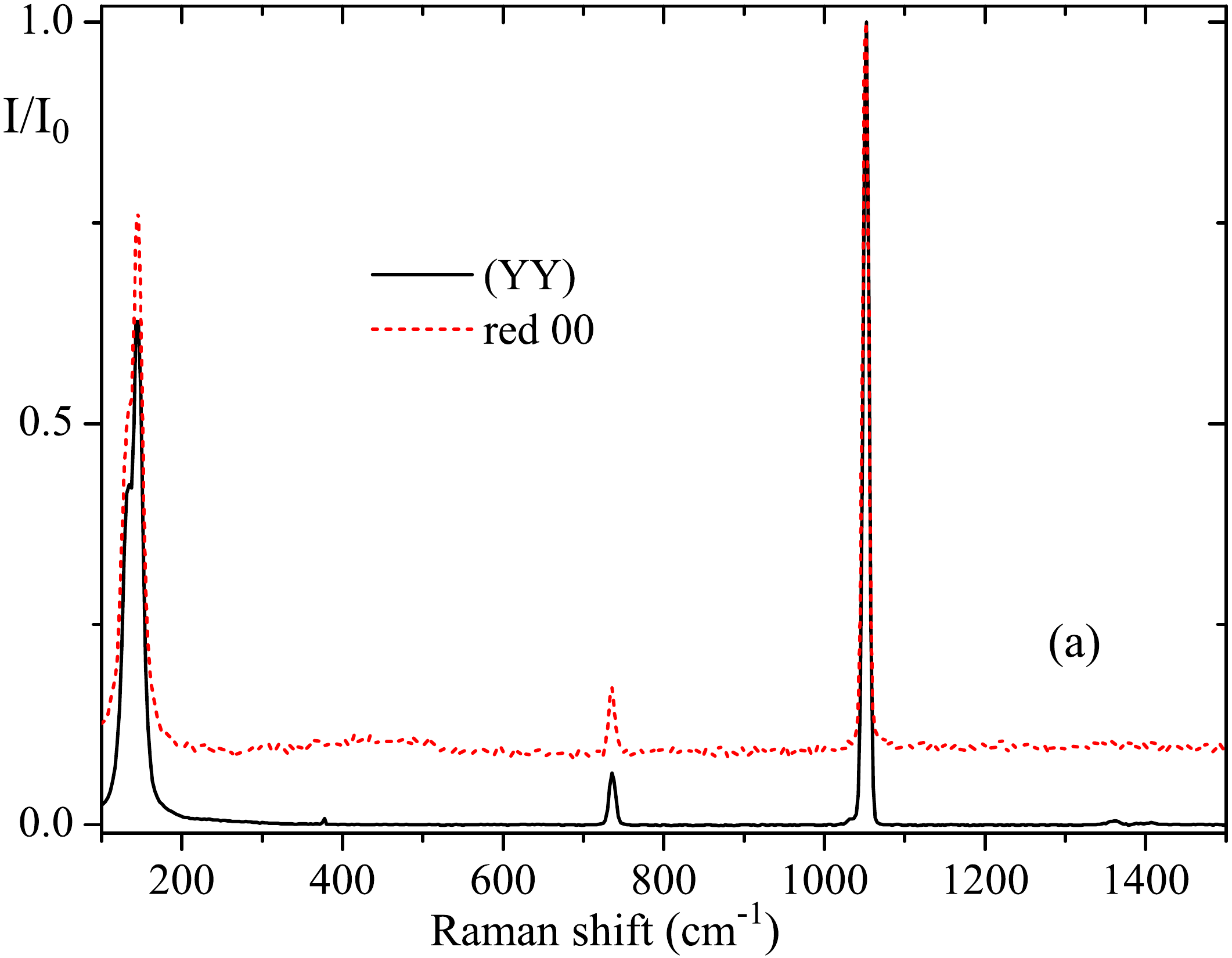}~ \includegraphics[scale=0.27]{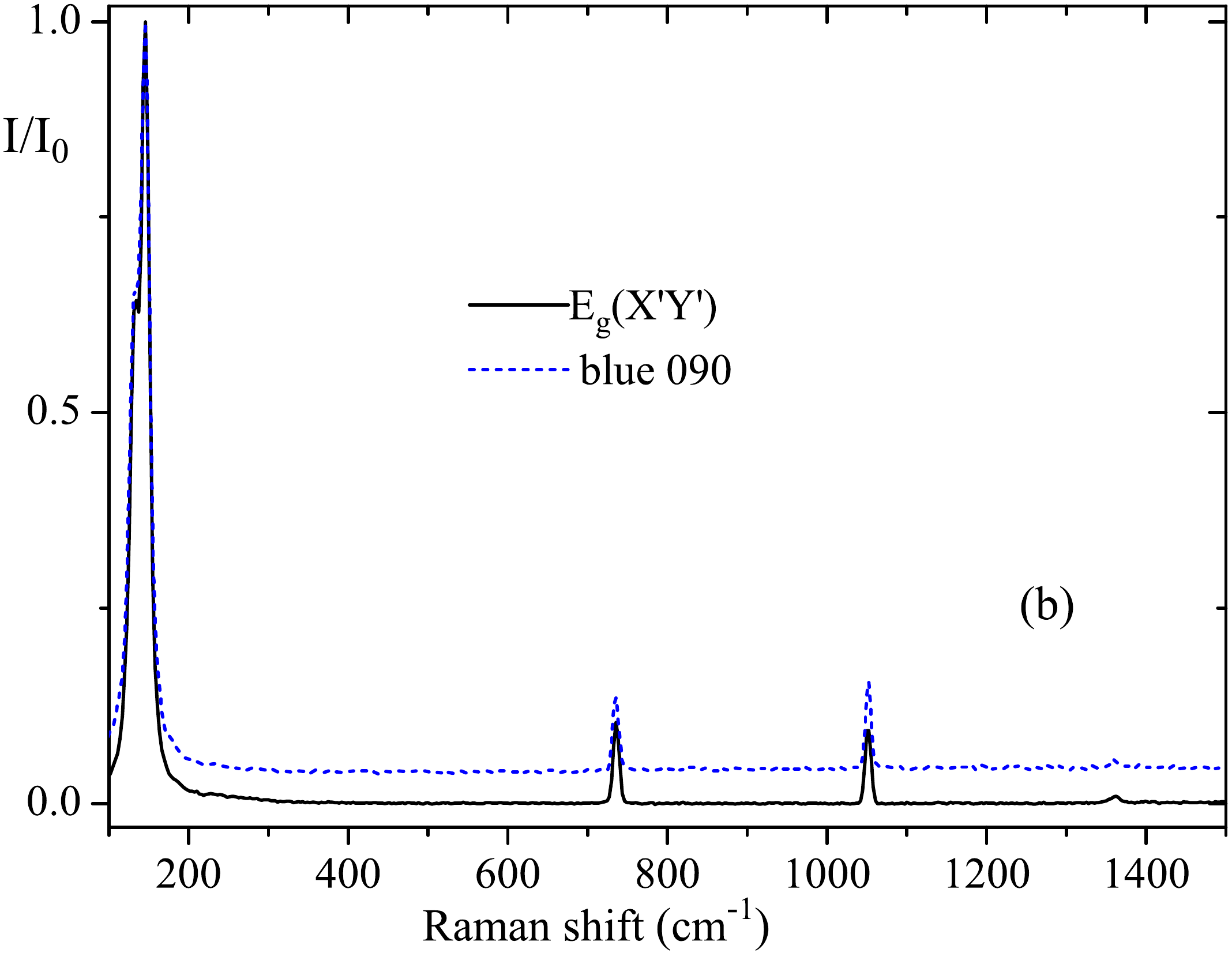}~
\includegraphics[scale=0.27]{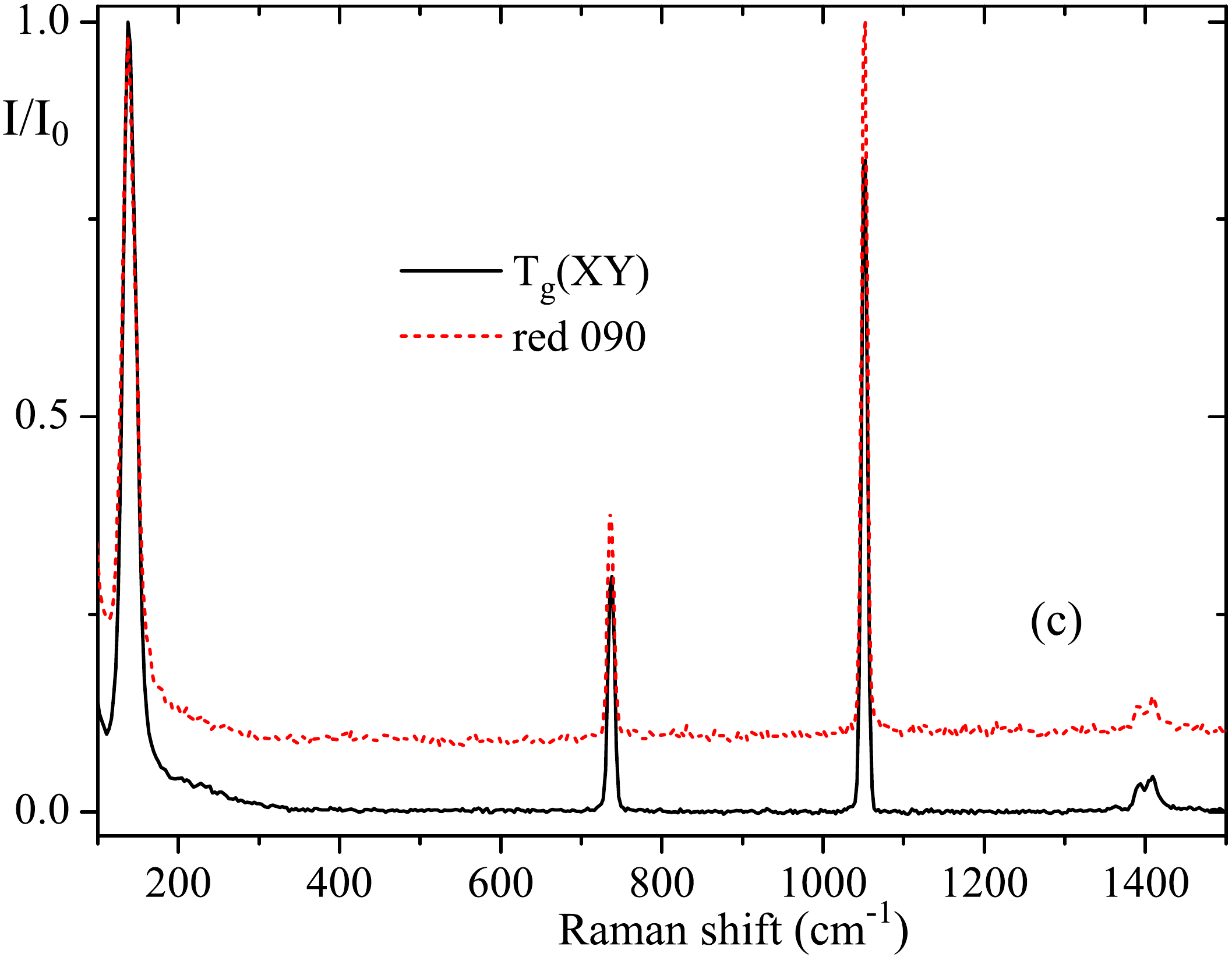}
\end{center}
\caption[]{Comparison of polarized Raman spectra taken from a single BNO crystal and from the pSiO$_2$:BNO composite. Black straight lines correspond to spectra recorded from a single BNO crystal, red/blue dashed lines present spectra reflected from the red/blue regions of SiO$_2$:BNO composite depicted in Figure \ref{Map}. All spectra are normalized to the highest intensity mode in each scattering geometry, namely to 1053 cm$^{-1}$ in (a) and (c), to 146 cm$^{-1}$ in (b).} \label{Ramnano}
\end{figure*}

Figure \ref{Ramnano} presents a comparison between some typical patterns of polarized Raman spectra recorded from comparatively large red and blue regions of nanocomposite shown in Figure \ref{Map} and the polarized spectra taken from the single BNO crystal. All spectra are normalized to the most intensive line which is the same for BNO and pSiO$_2$:BNO samples in the spectra depicted in Figure \ref{Ramnano} (a, b and c), respectively. As seen in this figure, there is a very close resemblance between the polarized Raman spectra taken from both the BNO single crystal and the pSiO$_2$:BNO composite. This similarity is especially prominent for the following pairs of spectra, XY and 090 red, X'Y' and 090 blue, YY and 00 red, respectively. Based on these Raman data one may conclude that the compound synthesized within the pSiO$_2$ pores is, with very high probability, a nanoscale BNO single crystal.

To analyze Raman spectra quantitatively we fitted the line shape of all spectra by using the Lorentz function for constituent spectral lines. A comparison of fitted parameters obtained for a BNO single crystal and for the pSiO$_2$:BNO composite is presented in Table \ref{Lorentz}. Note that the Raman data listed in "BNO YY" column correspond to spectra measured in single crystal using the Z(YY)-Z experimental geometry, \textit{i.e.} they contain the information about normal modes of both A$_g$ and E$_g$ species.

There is a very close correlation not only between peak positions of the single crystal and nanocomposite Raman spectra but between their peak intensities and FWHMs as well. Thereby one may infer that the regions indicated by red and blue color in Figure \ref{Map} mainly contain the single crystal parts of BNO crystal of $\sim$10-20 $\mu$m dimensions. This is a rather unexpected statement keeping in mind the non-homogeneous nature of $\sim$10-12 nm sized pores inherent for porous host pSiO$_2$ matrix and consequently the $\sim$10-12 nm dimensionality of BNO nanoscale crystals inserted in pSiO$_2$.

%%%%%%%%%%%%%%%%%%%%%%%%%%%%%%%%%%%%%%%%%%%%%%%%%%%%%%%%%%%%%%%%%%%%%%%%%%
%%%%%%%%%%%%%%%%%%%%%%%%%%%%%%%%%%%%%%%%%%%%%%%%%%
\begin{table} []
\caption{Parameters of Lorentzian fit of polarized Raman spectra recorded of BNO single crystal and of the red and blue regions of pSiO$_2$:BNO composite. Peak intensity was normalized to the strongest line in each spectrum. FWHM implies the full width at half maximum.}
\label{Lorentz}
\footnotesize
%\hspace*{-2em}
%\small
\begin{tabular}{cccccc}
\\\hline
\multicolumn{6}{c}{Peak position (cm$^{-1}$)}\\
\hline
BNO& composite&BNO& composite&BNO& composite\\
XY&090 red&X'Y'&090 blue&YY&00 red\\
\hline
137&138&132&132&132&131\\
147&148&146&146&145&145\\
737&737&736&735&736&735\\
1051&1051&1051&1052&1052&1051\\
1392&1391&1361&1360&&\\
1407&1409&&&&\\
\hline
\multicolumn{6}{c}{Normalized peak intensity}\\
\hline
BNO& composite&BNO& composite&BNO& composite\\
XY&090 red&X'Y'&090 blue&YY&00 red\\
\hline
0.90&0.81&0.54&0.54&0.29&0.29\\
0.41&0.14&1&1&0.55&0.60\\
0.35&0.30&0.12&0.11&0.06&0.08\\
1&1&0.12&0.14&1&1\\
0.03&0.03&0.01&0.01&&\\
0.05&0.04&&&&\\
\hline
\multicolumn{6}{c}{FWHM (cm$^{-1}$)}\\
\hline
BNO& composite&BNO& composite&BNO& composite\\
XY&090 red&X'Y'&090 blue&YY&00 red\\
\hline
14.4&17.1&13.5&15.0&12.7&16.3\\
12.4&11.1&11.7&12.3&12.7&12.9\\
7.5&7.0&7.1&7.39&7.4&7.7\\
6.4&6.2&6.4&6.77&6.7&6.6\\
10.0&9.5&13.0&9.80&&\\
16.3&15.0\\
\hline
\end{tabular}
\end{table}
%%%%%%%%%%%%%%%%%%%%%%%%%%%%%%%%%%%%%%%%%
It is quite instructive to look closer at the FWHMs of Raman lines since they may be influenced by the spatial confinement. An increase of the Raman line's FWHM with decrease of nanoscale grain size is a known phenomenon in Raman spectroscopy of nanomaterials \cite{Gouadec,Kosacki,Korepanov}. A confinement effect should have smaller impact on the high-frequency internal NO$_3$ vibrations owing to their strong covalent bonding. It is only slightly dependent on the crystal morphology, \textit{i.e.}, it is barely sensitive to the surrounding crystalline medium. The investigation of Raman spectra of ZnO nanoparticles of 4.6-12 nm size corroborates this assumption \cite{Korepanov}. ZnO is a crystal of wurtzite structure with a strong short-range zinc-oxygen covalent bonding. According to the paper \cite{Korepanov}, the Raman active high-frequency mode near $\sim$440 cm$^{-1}$ shows practically no nanograin size dependence within the 4.6-12 nm. The low-frequency mode of ZnO near $\sim$100 cm$^{-1}$ demonstrates a slight increase of Raman line's FWHM only for the smallest nanoparticles of 4.6 nm. However, the most significant size dependence manifest the localized acoustic phonon modes placed below 40 cm$^{-1}$ owing to the large frequency dispersion of acoustic modes \cite{Korepanov}.

Two types of crystal bonding, ionic and covalent, are inherent for BNO crystal. Strong covalent type of bonding is characteristic for the high-frequency internal vibrations of NO$_3$ groups, whereas the ionic crystal bonding is mainly typical for the low-frequency interactions between Ba$^{2+}$ and (NO$_3$)$^{-}$ ions. The comparatively weaker long-range ionic interactions should be more susceptible to distinctive changes of the internal crystal field resulting from the final crystal size and the interaction with the confining porous matrix. These considerations are corroborated by the data presented in Table \ref{Lorentz}. In the previous section we have shown by \textit{ab initio} calculation that the external lattice modes should not exceed 184 cm$^{-1}$, see Table \ref{Compar}. The FWHMs of low-frequency lattice modes of the pSiO$_2$:BNO composite located below $\sim$150 cm$^{-1}$ are larger, with one exception, than the FWHMs values of the macroscopic crystal, see the first and second rows in Table \ref{Lorentz}. For higher phonon frequencies this tendency is not observable. One may deduce that in the pSiO$_2$:BNO composite the spatial confinement effect has small, but quite distinctive impact on the FWHMs of lattice vibrations. However, almost no influence on the frequencies and peak intensities of phonon modes of nanoscale BNO crystals are detectable.

\section{Conclusions}
We have performed an exhaustive lattice dynamics study of BNO crystals embedded in a mesoporous pSiO$_2$ matrix. In order to accomplish the symmetry classification of phonon modes, the separation of A$_g$ and E$_g$ type vibrations has been carried out by the use of a special Raman scattering geometry. The rigorous group symmetry analysis of lattice dynamics and \textit{ab initio} calculation of phonon frequencies and \textit{eigen}vectors allowed us to interpret in detail the BNO Raman spectrum. Nevertheless, the \textit{ab initio} calculations performed within both GGA and GGA + DFT-D3 methods give much less LO-TO splitting of T$_u$ polar modes than found in the experimental data \cite{Bon}.

Based on the Raman scattering of the composite pSiO$_2$:BNO material one can conclude that BNO crystals of cubic symmetry form in pore space. The observation of dendritic single-crystal structures extending well beyond the size of the single nanopores indicates a propagation of the crystallization front through microporosity in the pore walls towards neighbouring pores. Despite the $\sim$10-12 nm pores of pSiO$_2$ matrix, the much larger single crystalline regions of $\sim$10-20 $\mu$m spacing of nearly uniform single BNO crystal appear within the pSiO$_2$ host compound. We have detected almost no impact of spatial confinement both on the position and the intensity of the Raman lines of the pSiO$_2$:BNO composite as compared to the bulk BNO single crystal. However, there is a small but recordable increase of FWHMs of low-frequency lattice vibrations located below $\sim$150 cm$^{-1}$. We explain this selective low-frequency impact on the FWHMs by a larger sensitivity of the ionic interatomic interactions to nanoconfinement in comparison to the strong covalent interactions dominating the high-frequency Raman response.

Overall, our study shows the power of combining Raman scattering, X-ray diffraction experiments with group-theory analysis and lattice dynamics calculation to study solution crystallization in monolithic nanoporous media. For the future we envisage analogous experiments on NaCl solutions and other mineral forming systems in confinement.

\section{CRediT authorship contribution statement}
Ya. Shchur: Conceptualization, Raman spectra analysis, symmetry analysis, first principles calculation, writing – original draft,; G. Beltramo: Raman scattering, methodology, data analysis; A.S. Andrushchak: funding acquisition, project administration; S. Vitusevich: data analysis, project
administration; P. Huber: X-ray analysis, sample preparation, writing – review$\&$editing; V. Adamiv: sample preparation; I. Teslyuk: sample preparation; N. Boichuk: sample preparation,data analysis; A.V. Kityk: methodology, X-ray analysis,  writing – review$\&$editing

\section{Declaration of Competing Interest}
The authors declare that they have no known competing financial
interests or personal relationships that could have appeared
to influence the work reported in this paper.

\section{Acknowledgement}
Ya. Shchur was supported by project No.6541030 of National Academy of Sciences of Ukraine. A.V. Kityk acknowledges a support from resources for science in years 2018-2022 granted for the realization of international co-financed project Nr W13/H2020/2018 (Dec. MNiSW 3871/H2020/2018/2). The presented results are part of a project that has received funding from the European Union Horizon 2020 research and innovation programme under the Marie Sklodowska-Curie grant agreement no. 778156. PH thanks for support by the Deutsche Forschungsgemeinschaft (DFG, German Research Foundation) Projektnummer 192346071, SFB 986 ''Tailor-Made Multi-Scale Materials Systems'', as well as by the Centre for Integrated Multiscale Materials Systems CIMMS, funded by Hamburg science authority.
Calculations have been carried out using resources provided by Wroclaw Centre for Networking and Supercomputing (http://wcss.pl), grant No. 160.

%\bibliography{KDP,HuberLabReferences}

\begin{thebibliography}{10}

\bibitem{Alba-Simionesco2006}
C~Alba-Simionesco, B~Coasne, G~Dosseh, G~Dudziak, K~E Gubbins, R~Radhakrishnan,
  and M~Sliwinska-Bartkowiak.
\newblock {Effects of confinement on freezing and melting}.
\newblock {\em J. Phys.: Condens. Matter}, 18:R15, 2006.

\bibitem{Desarnaud2014}
Julie Desarnaud, Hannelore Derluyn, Jan Carmeliet, Daniel Bonn, and Noushine
  Shahidzadeh.
\newblock {Metastability Limit for the Nucleation of NaCl Crystals in
  Confinement}.
\newblock {\em The Journal of Physical Chemistry Letters}, 5(5):890--895, 3
  2014.

\bibitem{Huber2015}
Patrick Huber.
\newblock {S}oft matter in hard confinement: phase transition thermodynamics,
  structure, texture, diffusion and flow in nanoporous media (invited topical
  review).
\newblock {\em Journal of Physics: Condensed Matter}, 27(10):103102, 2015.

\bibitem{Lindstrom2016}
Nadine Lindstr{\"{o}}m, Tanya Talreja, Kirsten Linnow, Amelie Stahlbuhk, and
  Michael Steiger.
\newblock {Crystallization behavior of Na2SO4–MgSO4 salt mixtures in
  sandstone and comparison to single salt behavior}.
\newblock {\em Applied Geochemistry}, 69:50--70, 6 2016.

\bibitem{Meldrum2020}
Fiona~C. Meldrum and Cedrick O'Shaughnessy.
\newblock {Crystallization in Confinement}.
\newblock {\em Advanced Materials}, 32(31):2001068, 8 2020.

\bibitem{Wallacher2002}
D~Wallacher, V~P Soprunyuk, A~V Kityk, and K~Knorr.
\newblock {Dielectric response of CO and Ar condensed into mesoporous glass RID
  B-8351-2008 RID B-8142-2008}.
\newblock {\em Physical Review B}, 66(1):14203, 7 2002.

\bibitem{Christenson2001}
H~K Christenson.
\newblock {Confinement effects on freezing and melting}.
\newblock {\em Journal of Physics-condensed Matter}, 13(11):R95--R133, 3 2001.

\bibitem{Huber2004}
P~Huber, D~Wallacher, J~Albers, and K~Knorr.
\newblock {Quenching of lamellar ordering in an n-alkane embedded in
  nanopores}.
\newblock {\em Europhys. Lett.}, 65:351, 2004.

\bibitem{Zeng2018}
Muling Zeng, Yi-Yeoun Kim, Clara Anduix-Canto, Carlos Frontera, David Laundy,
  Nikil Kapur, Hugo~K. Christenson, and Fiona~C. Meldrum.
\newblock {Confinement generates single-crystal aragonite rods at room
  temperature}.
\newblock {\em Proceedings of the National Academy of Sciences},
  115(30):7670--7675, 7 2018.

\bibitem{Sentker2018}
Kathrin Sentker, Arne~W. Zantop, Milena Lippmann, Tommy Hofmann, Oliver~H.
  Seeck, Andriy~V. Kityk, Arda Yildirim, A.~Sch{\"o}nhals, Marco~G. Mazza, and
  Patrick Huber.
\newblock {Q}uantized {S}elf-{A}ssembly of {D}iscotic {R}ings in a {L}iquid
  {C}rystal {C}onfined in {N}anopores.
\newblock {\em Phys. Rev. Lett.}, 120(6):067801, 2018.

\bibitem{Meissner2019}
Jens Meissner, Albert Prause, and Gerhard~H. Findenegg.
\newblock {Secondary Confinement of Water Observed in Eutectic Melting of
  Aqueous Salt Systems in Nanopores}.
\newblock {\em The Journal of Physical Chemistry Letters}, 7(10):1816--1820, 5
  2016.

\bibitem{Enninful2021}
Henry R N~B Enninful, Daniel Schneider, Dirk Enke, and Rustem Valiullin.
\newblock {Impact of Geometrical Disorder on Phase Equilibria of Fluids and
  Solids Confined in Mesoporous Materials}.
\newblock {\em LANGMUIR}, 37(12):3521--3537, 2021.

\bibitem{Sentker2019}
Kathrin Sentker, Arda Yildirim, Milena Lippmann, Arne Zantop, Florian Bertram,
  Tommy Hofmann, Oliver~H Seeck, Andriy~V. Kityk, Marco~G. Mazza, Andreas
  Sch{\"o}nhals, and Patrick Huber.
\newblock Self-assembly of liquid crystals in nanoporous solids for adaptive
  photonic metamaterials.
\newblock {\em Nanoscale}, 11(48):23304--23317, 2019.

\bibitem{Shi}
G.~Shi and E.~Kioupakis.
\newblock {Electronic and optical properties of nanoporous silicon for
  solar-cell applications}.
\newblock {\em ACS Photonics}, 2:208--215, 2015.

\bibitem{LHP}
Ya. Shchur and A.V. Kityk.
\newblock {Ordered PbHPO$_4$ nanowires: Crystal structure, energy bands and
  optical properties from first principles}.
\newblock {\em Computation. Materials Science}, 138:1--9, 2017.

\bibitem{Tang}
J.~Tang, H.-T. Wang, D.~H. Lee, M.~Fardy, Z.~Huo, T.P. Russel, and P.~Yang.
\newblock {Holey silicon as an efficient thermoelectric material}.
\newblock {\em Nano Lett.}, 10:4279--4283, 2010.

\bibitem{Hosseini}
S.A. Hosseini, G.~Romano, and P.A. Greaney.
\newblock {Mitigating the effect of nanoscale porosity on thermoelectric power
  factor of Si}.
\newblock {\em ACS Appl. Energy Mater.}, 4:1915--1923, 2021.

\bibitem{porSi}
Ya. Shchur, O.~Pavlyuk, A.S. Andrushchak, S.~Vitusevich, and A.V. Kityk.
\newblock {Porous Si partially filled with water molecules - crystal structure,
  energy bands and optical properties from first principles}.
\newblock {\em Nanomaterials}, 10:396:1--17, 2020.

\bibitem{Dasi}
G.~Dasi, T.~Lavanya, S.~Suneetha, S.~Vijayakumar, J.~Shim, and K.~Thangaraju.
\newblock {Raman and X-ray photoelectron spectroscopic investigation of
  solution processed Alq3/ZnO hybrid thin films}.
\newblock {\em Spectrochim. Acta A}, 265:120377, 2022.

\bibitem{Sharikova}
A.~Sharikova, Z.I. Foraida, L.~Sfakis, L.~Peerzada, M.~Larsen, J.~Castracane,
  and A.~Khmaladze.
\newblock {Characterization of nanofibers for tissue engineering: Chemical
  mapping by Confocal Raman microscopy}.
\newblock {\em Spectrochim. Acta A}, 227:117670, 2020.

\bibitem{Dhara}
D.~Das, P.~Dadhich, P.~Pal, S.~Thakur, S.~Neogi, and S.~Dhara.
\newblock {Carbon nano dot decorated copper nanowires for SERS-Fluorescence
  dual-mode imaging/anti-microbial activity and enhanced angiogenic activity}.
\newblock {\em Spectrochim. Acta A}, 227:117669, 2020.

\bibitem{Trajic}
J.~Trajic, R.~Kostic, N.~Romcevic, M.~Mitric, V.~Lazovic, P.~Balaz, and
  D.~Stojanovic.
\newblock {Raman spectroscopy of ZnS quantum dots}.
\newblock {\em J. Alloys Compd.}, 637:401--406, 2015.

\bibitem{Krajczewski}
J.~Krajczewski, A.~Michałowska, and A.~Kudelski.
\newblock {Star-shaped plasmonic nanostructures: New, simply synthetized
  materials for Raman analysis of surfaces}.
\newblock {\em Spectrochim. Acta A}, 225:117469, 2020.

\bibitem{Shchur_KDP}
Ya. Shchur, A.V. Kityk, V.V. Strelchuk, A.S. Nikolenko, N.A. Andrushchak,
  P.~Huber, and A.S. Andrushchak.
\newblock {Paraelectric KH$_2$PO$_4$ nanocrystals in monolithic mesoporous
  silica: Structure and lattice dynamics}.
\newblock {\em J. Alloys Compd.}, 868:159177:1--8, 2021.

\bibitem{Karpukhin}
S.N. Karpukhin and A.I. Stepanov.
\newblock {Generation of radiation in a resonator under conditions of
  stimulated Raman scattering in Ba(NO$_3$)$_2$, NaNO$_3$, and CaCO$_3$
  crystals}.
\newblock {\em Sov. J. of Quantum Electronics}, 16(8):1027--1031, 1986.

\bibitem{Murray}
J.T. Murray, R.C. Powell, N.~Peyghambarian, D.~Smith, W.~Austin, and R.A.
  Stolzenberger.
\newblock {Generation of 1.5-mm radiation through intracavity solid-state Raman
  shifting in Ba(NO$_3$)$_2$ nonlinear crystals}.
\newblock {\em Optics Letters}, 20(9):1017--1019, 1995.

\bibitem{Zverev}
P.G. Zverev, T.T. Basiev, and A.M. Prokhorov.
\newblock {Stimulated Raman scattering of laser radiation in Raman crystals}.
\newblock {\em Optical Mater.}, 11:335--352, 1999.

\bibitem{Eremenko}
A.S. Eremenko, S.N. Karpukhin, and A.I. Stepanov.
\newblock {Stimulated Raman scattering of the second harmonic of a neodymium
  laser in nitrate crystals}.
\newblock {\em Sov. J. of Quantum Electronics}, 10(1):113--114, 1980.

\bibitem{Perdew}
J.P. Perdew and Y.~Wang.
\newblock {Accurate and simple analytic representation of the electron-gas
  correlation energy}.
\newblock {\em Phys. Rev. B}, 45:13244: 1--6, 1992.

\bibitem{Grimme}
S.~Grimme, J.~Antony, S.~Ehrlich, and H.~Krieg.
\newblock {A consistent and accurate ab initio parametrization of density
  functional dispersion correction (DFT-D) for the 94 elements H-Pu}.
\newblock {\em J. Chem. Phys.}, 132:13244: 1--2, 2010.

\bibitem{Abinit1}
X.~Gonze, G.-M. Rignanese, M.~Verstraete, J.-M. Beuken, Y.~Pouillon,
  R.~Caracas, F.~Jollet, M.~Torrent, G.~Zerah, M.~Mikama, P.~Ghosez,
  M.~Veithen, J.Y. Raty, V.~Olevano, F.~Bruneval, L.~Reining, R.~Godby,
  G.~Onida, D.R. Hamann, and D.C. Allan.
\newblock A brief introduction to the {ABINIT} software package.
\newblock {\em Z. Kristallogr.}, 220:558: 1--5, 2005.

\bibitem{Abinit2}
X.~Gonze, B.~Amadon, P.-M. Anglade, J.-M. Beuken, F.~Bottin, P.~Boulanger,
  F.~Bruneval, D.~Caliste, R.~Caracas, M.~Cote, T.~Deutsch, L.~Genovese, Ph.
  Ghosez, M.~Giantomassi, S.~Goedecker, D.R. Hamann, P.~Hermet, F.~Jollet,
  G.~Jomard, S.~Leroux, M.~Mancini, S.~Mazevet, M.J.T. Oliveira, G.~Onida,
  Y.~Pouillon, T.~Rangel, G.-M. Rignanese, D.~Sangalli, R.~Shaltaf, M.~Torrent,
  M.J. Verstraete, G.~Zerah, and J.W. Zwanziger.
\newblock {ABINIT: First-principles approach to material and nanosystem
  properties}.
\newblock {\em Comput. Phys. Commun}, 180:2582: 1--38, 2009.

\bibitem{Monkhorst}
H.J. Monkhorst and J.D. Pack.
\newblock {Special points for Brillouin-zone integrations}.
\newblock {\em Phys. Rev. B}, 13:5188--5191, 1976.

\bibitem{Nowotny}
H.~Nowotny and G.~Heger.
\newblock {Structure refinement of strontium nitrate, Sr(NO$_3$)$_2$, and
  barium nitrate, Ba(NO$_3$)$_2$}.
\newblock {\em Acta Crystallogr. C}, 39:952--956, 1983.

\bibitem{Broyden}
C.G. Broyden.
\newblock {The convergence of a class of double-rank minimization algorithms:
  General considerations}.
\newblock {\em J. Inst. Maths. Appl.}, 6:76: 1--15, 1970.

\bibitem{Patterson}
A.L. Patterson.
\newblock {The Scherrer formula for X-ray particle size determination}.
\newblock {\em Phys. Rev.}, 56:978--982, 1939.

\bibitem{Waszkowska2021SHG}
Karolina Waszkowska, Pierre Josse, Clément~Clement Cabanetos, Philippe
  Blanchard, Bouchta Sahraoui, Dominique Guichaoua, Igor Syvorotka, Olha Kityk,
  Robert Wielgosz, Patrick Huber, and Andriy~V. Kityk.
\newblock {Anisotropic confinement of chromophores induces second-order
  nonlinear optics in a nanoporous photonic metamaterial}.
\newblock {\em Optics Letters}, 46(4):845, 2 2021.

\bibitem{Poulet}
H.~Poulet and J.P. Mathieu.
\newblock {\em {Vibration Spectra and Symmetry of Crystals}}.
\newblock Gordon and Breach, 1976.

\bibitem{Nakamoto}
K.~Nakamoto.
\newblock {\em Infrared and Raman spectra of inorganic and coordination
  compounds}.
\newblock John Wiley \& Son, 1997.

\bibitem{Lockwood}
D.J. Lockwood.
\newblock {Isolating the totally symmetric Raman spectrum of cubic crystals:
  the A$_1$ spectrum of Cr$_3$B$_7$O$_{13}$Cl}.
\newblock {\em J. Raman Spectrosc.}, 2:555--562, 1974.

\bibitem{Bon}
A.M. Bon, C.~Benoit, and O.~Bernard.
\newblock {Dynamical properties of crystals of Sr(NO$_3$)$_2$, Ba(NO$_3$)$_2$
  and Pb(NO$_3$)$_2$. I Infrared spectra and structure}.
\newblock {\em Phys. Status Solidi B}, 78:67--78, 1976.

\bibitem{CDP2}
Ya. Shchur.
\newblock {On the vibrational properties of CsD$_2$PO$_4$ crystal}.
\newblock {\em Phys. Status Solidi B}, 244(2):569--577, 2007.

\bibitem{DRDP}
Ya. Shchur.
\newblock {On the issue of superstructure phase transitions in monoclinic
  RbD$_2$PO$_4$ crystal}.
\newblock {\em J. Phys.: Condens. Matter}, 20:195212: 1--8, 2008.

\bibitem{TDP}
Ya. Shchur.
\newblock {Phase transitions in TlH$_2$PO$_4$ and TlD$_2$PO$_4$ crystals:
  lattice dynamical treatment}.
\newblock {\em J. Phys.: Condens. Matter}, 22:315902: 1--8, 2010.

\bibitem{LHP2}
Ya. Shchur.
\newblock {Vibrational analysis of PbHPO$_4$ and PbDPO$_4$ crystals}.
\newblock {\em Phys. Status Solidi B}, 246(1):102--109, 2009.

\bibitem{BTBO}
D.~Kasprowicz, T.~Zhezhera, A.~Lapinski, M.~Chrunik, A.~Majchrowski, A.V.
  Kityk, and Ya. Shchur.
\newblock {Lattice dynamics of Bi$_3$TeBO$_9$ microcrystals: micro-Raman/IR
  spectroscopic investigation and \textit{ab initio} analysis}.
\newblock {\em J. Alloys Compd.}, 782:488--495, 2019.

\bibitem{Kondrashova2017}
Daria Kondrashova, Alexander Lauerer, Dirk Mehlhorn, Hervé Jobic, Armin
  Feldhoff, Matthias Thommes, Dipanjan Chakraborty, Cedric Gommes, Jovana
  Zecevic, Petra de~Jongh, Armin Bunde, Jörg K{\"{a}}rger, and Rustem
  Valiullin.
\newblock {Scale-dependent diffusion anisotropy in nanoporous silicon}.
\newblock {\em Scientific Reports}, 7(1):40207, 2 2017.

\bibitem{Gouadec}
G.~Gouadec and P.~Colomban.
\newblock {Raman Spectroscopy of nanomaterials: How spectra relate to disorder,
  particle size and mechanical properties}.
\newblock {\em Prog. Cryst. Growth Charact. Mater.}, 53:1--56, 2007.

\bibitem{Kosacki}
I.~Kosacki, T.~Suzuki, H.U. Anderson, and P.~Colomban.
\newblock {Raman scattering and lattice defects in nanocrystalline CeO$_2$ thin
  films}.
\newblock {\em Solid State Ion.}, 149:99--105, 2002.

\bibitem{Korepanov}
V.I. Korepanov, S.-Y. Chan, H.-C. Hsu, and H.-o~A. Hamaguchi.
\newblock {Phonon confinement and size effect in Raman spectra of ZnO
  nanoparticles}.
\newblock {\em Heliyon}, 5:e01222--1--14, 2019.

\end{thebibliography}
%\bibliographystyle{unsrt}
%%%%%%%%%%%%%%%%%%%%%%%%%%%%%%%%%%%

%%%%%%%%%%%%%%%%%%%%%%%%%%%%%%%
\end{document}